\documentclass{aa501}
\usepackage{graphics,epsf,epsfig,graphicx}

\begin{document}

\title{Gas accretion on spiral galaxies:
\\ bar formation and renewal}

\author{F. Bournaud \inst{1} \and  F. Combes \inst{2}} 
\offprints{F. Bournaud, \email{bournaud@clipper.ens.fr}} 
\institute{
Ecole Normale Sup\'erieure, 45 rue d'Ulm, F-75005, Paris, France
\and
Observatoire de Paris, DEMIRM, 61 Av. de l'Observatoire, 
F-75014, Paris, France 
} 
\date{Received XX XX, 2002; accepted XX XX, 2002}
\authorrunning{Bournaud \& Combes} 
\titlerunning{}

\abstract{ The effects of gas accretion on spiral
disk dynamics and stability are studied through N-body
simulations, including star formation and gas/stars
mass exchange.
The detailed processes of bar formation,
bar destruction and bar re-formation are followed,
while in the same time the disk to bulge ratio is
varying. The accreted gas might be first prevented
to flow inwards to the center by the bar gravity torques,
which maintains it to the outer Lindblad resonance. While 
the first bar is weakening, the accreted gas replenishes
the disk, increasing the disk-to-bulge ratio, and
the disk self-gravity. A second bar is then unstable,
with a higher pattern speed, due both to the increased
mass, and shorter bar length. Three or four bar episodes
have been followed over a Hubble time. Their strength
is decreasing with time, while their pattern speed is
increasing. Detailed balance of the angular momentum
transfer and evolution can account for these processes.
 The gas recycled through star formation, 
and rejected through stellar mass loss plays also a 
role in the disk dynamics.
Implications on the spiral galaxy dynamics and evolution
along the Hubble sequence, and as a function of redshift
are discussed.
}

\maketitle

\section{Introduction} 
Bars are an essential feature in galaxy evolution. About two-thirds of
spiral galaxies are barred (de Vaucouleurs 1963), one third being strongly
barred (SB), the other third mildly barred (SAB). In the near-infrared,
where the central structure is unveiled from dust, the fraction
is even higher (Eskridge et al. 2000). 

In the last decade, it has been realized that bars are not long lived
features, in particular in gaseous spiral disks, and that galaxies are
not frozen for a Hubble time in their morphological class:
a bar can be destroyed by large radial gas inflow and mass accumulation
in the center (Hasan \& Norman 1990, Pfenniger \& Norman 1990, 
Hasan et al. 1993, Friedli \& Benz 1993). The mass concentration destroys 
the orbital structure that supported the bar, and this begins through the
creation of two strong inner Lindblad resonances (ILR).
The process is initiated by the strong gravity torques exerted
by the bar on the gaseous spiral arms, which decrease afterwards
as the bar weakens. It can lead to the decoupling of 
a secondary bar, embedded into the primary, just inside its ILR
(Friedli \& Martinet 1993, Combes 1994), and this is likely to 
weaken the primary bar. 

The bar phenomenon can then be a self-regulated process (e.g. 
Combes 2000). A tantalizing scenario is that a galaxy may have several 
bar episodes in its life, and that the fraction of barred galaxies
in the de Vaucouleurs classification only reflects the time percentage
that a spiral galaxy spends as a barred object. How can a bar
be revived after a destruction or weakening event? The disk
self-gravity must be enhanced to counteract the stabilizing
influence of the bulge and central mass concentration, enforced
by the first bar episode. This may be obtained through external
gas accretion, that settles in the disk. Significant amounts of matter 
can be accreted in the life-time of the galaxy (e.g. Katz et al. 1996),
and we consider realistic that the mass of a galaxy could double
in a few Gyrs. The presence of gaseous warps in practically every 
spiral galaxy (Sancisi 1983, Briggs 1990) has been interpreted as a sign that a
spiral galaxy accretes mass and angular momentum: the amount accreted
is such that it changes completely orientation in a typical time-scale of a 
few Gyrs (Jiang \& Binney, 1999).

During its life, a bar changes its pattern speed, and this could be
a possible way to trace its age. In dissipationless simulations,
the pattern speed decreases in a time-scale of a few Gyrs 
(Combes \& Sanders 1981). This is also related to the length of
the bar. The early bar instability involves central orbits, where 
the orbital precession rate is higher, then more and more orbits
become trapped in the bar, at larger radii, and the bar lengthens
and slows down, The transient spiral waves formed in the stellar 
component transfer the angular momentum outwards. This process
is also accelerated by escaping chaotic orbits, around corotation
(Pfenniger \& Friedli 1991). 
If there is a massive spheroidal dark matter halo, concentrated
enough to perturb the bar dynamics, the pattern speed of the bar
could decrease even further, owing to the dynamical friction
(Debattista \& Sellwood 1998). According to the density
of dark matter, the time scale could be estimated down to a few
10$^8$ yr. Since bars are observed most of the time to be
fast rotators, this puts constraints on the density of
dark matter in the central parts of galaxies. Fluctuations of the 
pattern speed can also be triggered by interacting companions
(Gerin et al. 1990, Miwa \& Noguchi 1998), but these remain 
transient.
If gas accretion is able to rejuvenate the bars, it is important
to know the implications on the pattern speed, that could 
be a tracer of the event. This is studied in detail in the 
presented simulations.

In Sect.~2 we briefly present the code and numerical methods used. In Sects.~3, 
4 and 5 we present the results from our simulations. 
We give our conclusions in Sect.~6.

\section{Physical model for disk galaxies and numerical techniques}

To explore the influence of gas accretion on the disk dynamics,
there are many free parameters to vary; in a first step, to 
increase computing efficiency, we consider only 2-D models,
with high spatial resolution and low CPU cost, with the constraint 
that the accretion has the same angular momentum direction as
the initial galaxy disk. In a second step, with 3-D models, the
angular momentum of the accreted matter is varied.

We first describe the two-dimensional models and present the associated 
results in Sect.~3. We present the three-dimensional models, 
with a lower resolution, in Sect.~4.

        \subsection{Disk, bulge and halo}
The galactic disk is first considered as an infinitely flattened
 two-dimensional component, embedded in a 
spherical halo, and a bulge partly stabilizes its inner regions. The initial 
disk is a Toomre disk, with a surface density of the form (Toomre, 1963):
\begin{equation} \label{Toomredisk}
\mu(r)=\mu_0 \left( 1+\frac{r^2}{a_{disk}^2} \right)^{-\frac{1}{2}}
\end{equation}
The scale-length $a_{disk}$ has been varied between 4 and 6 kpc.
The disk is initially truncated at a radius $r_{disk}$ = 15 kpc. 
This disk is made up of stellar and gaseous particles.
Initially, the number of particles 
is $2\cdot 10^5$, the disk mass is 
M$_{disk}^{initial}=4\cdot 10^{10}\;\mathrm{M}_{\sun}$, and its gas mass 
fraction is 30\%. The disk mass M$_{disk}$ and the number of gaseous particles 
increase with time because of gas accretion, as described in Sect.~\ref{accreting}.

In 2-D, the composition and the evolution of the bulge and the halo are not  
studied. They are only represented as analytical potentials. They are 
implemented in our simulations as Plummer spheres. The associated 
potentials respectively read:
\begin{equation} \label{phib}
\phi_{bulge}(r) = G \cdot M_B \left( 1+\frac{r^2}{a_B^2} \right)^{-\frac{1}{2}}
\end{equation}
and
\begin{equation} \label{phih}
\phi_{halo}(r) = G \cdot M_H \left( 1+\frac{r^2}{a_H^2} \right)^{-\frac{1}{2}}
\end{equation}
The bulge and halo masses, $M_B$ and $M_H$, are varied according 
to table \ref{param1}.
We here just note that $M_B$ is generally smaller than $M_{disk}$, and 
that the halo mass included in the disk radius, $M_{halo}^{inner}$, which 
is only a fraction of $M_H$, is typically at most of the same order as 
$M_{disk}$. The $M_B / M_{disk}^{initial}$ and $M_H / M_{disk}^{initial}$ ratios 
are the two main parameters. On the other hand, the bulge and halo 
scale-lengths, $a_B$ and $a_H$, are fixed at $a_B=1.4$ kpc and $a_H=25$ 
kpc (excepted in Sect.~\ref{arms}).

        \subsection{Gravitational interaction and gas dissipation}
The gravitational potential is computed via Fourier transforms. The 
particles are meshed on a useful Cartesian grid of $512\times512$  cells, and 
each particle mass is assigned to four grid points according to the 
"cloud-in-cell" interpolation. Fourier transforms, carried out on a
$1024 \times 1024$ grid in order to prevent Fourier periodic images do
disturb the real disk, allow us to calculate 
the disk own potential, and derive forces.
Then the analytical forces of the bulge and 
the halo are added, as they are described by (\ref{phib}) and (\ref{phih}). The 
potential differentiation and a bilinear interpolation give the gravitational 
force exerted on each particle. To suppress two-body relaxation, 
inefficient in real galaxies due to the large number of particles, 
the gravitational forces are softened with a Plummer softening,
of characteristic size of 200pc.

To represent the dissipative character of the gas component,
a frequently used method is the sticky-particles algorithm,
which schematizes the cloud-cloud collisions, and the non-elasticity
of the collisions, where the relative kinetic energy is lost
to heating the gas, which efficiently re-radiates it away
(e.g. Combes \& Gerin 1985). The whole process tends to
reduce the relative velocity of neighboring gas particles,
cooling the gas through reducing its velocity dispersion. 
 Here the process is represented by a dissipative force: 
\begin{equation}
\vec{f}=-k (\vec{v}-<\vec{v}>)
\end{equation}
The local mean speed $<\vec{v}>$ is computed on a $1024\times1024$ grid. 
The dissipation time-scale $k^{-1}$ is chosen between 5 and 10 Gyr.

The code finally computes the integration of equations of motion 
with a leap-frog algorithm (Hut et al. 1995). The time step for the gravity
potential computation, and for the integration of equations of
motion, $\Delta t _{\phi}$, varies with 
time and is chosen to satisfy:

\begin{equation}
\Delta t _{\phi}<\frac{L_{cell}}{V_{max}}
\end{equation}
where $L_{cell}$ is the cell size, and $V_{max}$ the maximum of the 
particles velocities. Computing $V_{max}$ allows us to make 
$\Delta t _{\phi}$ vary proportionally to $L_{cell} / V_{max}$. 
The time step then varies between 0.4 and 1. Myr. Then the effective 
number of times steps to reach the Hubble time is about 20000.

        \subsection{Initial velocities}
While creating the initial disk, with a spatial distribution
 as described by (\ref{Toomredisk}), the code 
computes the mean circular velocity in each cell, the rotational
frequency $\Omega$, the epicyclic frequency $\kappa$, etc..
Random motions are given to all particles to prevent 
axisymmetric instabilities, their magnitude being at least
equal to the critical speed (Toomre, 1964). The Toomre parameter $Q$
is typically $Q=1.5$.
The epicyclic approximation is used to relate the azimuthal and
radial velocity dispersions:
\begin{equation}
\sigma_\theta = \frac{\kappa}{2 \Omega} \sigma_r
\end{equation}
The rotational velocity is reduced from the circular speed, 
according to the Jeans equations, applied to the infinitely 
thin disk.

        \subsection{Star formation and stellar mass-loss}
The star formation scheme and the way the stars lose mass to the gas
are inspired from those described by Jungwiert et al. (2001). 
The stellar mass-loss is not approximated as instantaneous,
but decreases with time as a power-law.

The star formation is assumed to be described by a local
Schmidt law (Schmidt, 1959), i.e.
the star formation rate is proportional to a power of
the surface density of gas $\mu_{gas}^N$.
 We choose $N=1.4$, according to the observational determination 
of this parameter, on a global way and over
large scales (see Kennicutt, 1998a,b). The gas density is computed 
on a $1024\times1024$ grid, then the star formation rate is 
calculated for each cell of this grid. In each cell, the mass of 
young stars is taken away from all gas particles in the cell, in 
proportion to their mass at this epoch. This new star mass is 
added to all stellar particles present in the cell, in proportion 
to their mass. If there is no stellar particle in the cell, we 
wait a few time steps, until such a particle comes into the cell. 
This delays the actual star formation, yet, for 95$\%$ of the disk 
surface (the whole disk excepted its outer edge), such a situation 
occurs less than once 20 time steps: so, the delay in the star 
formation is generally avoided. When some mass is redistributed 
from a gas particle to a stellar one, the associated linear momentum is 
redistributed, too. When a mass $\Delta m$ is transferred from particle 
$1$ (gas) to particle $2$ (stellar), with initial masses $m_1$ and 
$m_2$ and initial velocities $\vec{v}_1$ and $\vec{v}_2$, the masses 
and velocities of particles are changed to:
\begin{equation}
m'_1=m_1-\Delta m
\end{equation}
\begin{equation}
m'_2=m_2+\Delta m
\end{equation}
\begin{equation}
\vec{v}'_1=\vec{v}_1
\end{equation}
\begin{equation}
\vec{v}'_2=\frac{m_2 \vec{v}_2+ \Delta m \vec{v}_1}{m'_2}
\end{equation}

This choice is not the only solution: the gas velocity is here
conserved during star formation. One might also choose to add 
some velocity dispersion to new stars.

The number of particles is then conserved, but the particle masses
vary. When a particle is accreted, it has the same mass as initially
present gaseous and stellar particles. The mass of stellar particles
doubles in about 15 Gyr, while gaseous particles particles
lose about half of their initial mass in 10 Gyr.

Since each particle represents in fact already a cluster of
actual components (star or gas cloud), the minimum mass of each particle
being of the order of 10$^6$ M$_\odot$, the stellar mass-loss
represents for each particle, the average mass-loss of an
ensemble of stars, of various masses. As described in Jungwiert 
et al. (2001), we assume that a coeval stellar population has a mass-loss 
rate $\dot{M}_{mass-loss}(t)$ equal to:
\begin{equation} \label{coeval}
\dot{M}_{mass-loss}(t)=M_{initial} \frac{c_0}{t-t_b+T_0}
\end{equation}
where $M_{initial}$ is the initial mass of the stellar population 
studied, and $t_b$ its birth time. The two parameters 
are fixed to $c_0=5.35\cdot10^{-2}$ and $T_0=4.86$ Myr.

However, such a mass-loss description is suitable only for a population of stars 
born in the same stellar cluster. We may use it for the initial particles, but it 
is not convenient for particles containing stars born at different epochs. Yet we 
have to add the mass of young stars in particles 
containing older ones in the above star 
formation scheme. Such additions may change the mass-loss rate of particles. Thus, we 
will define the mass-loss law of any particle in a recursive way:
\begin{itemize}
\item At $t=0$, any particle has a mass-loss rate given by (\ref{coeval}).
\item When new stars mass is added to a particle, we consider them as two 
merging particles: the old and young stellar particles. The 
first one is supposed to have a mass-loss rate given by:
\begin{equation}
\dot{M}_{old}(t)=\frac{c_{old}}{t+T_{old}}
\end{equation}
The second one contains a coeval population, therefore it has a mass-loss rate 
described by (\ref{coeval}):
\begin{equation}
\dot{M}_{young}(t)=\frac{c_{young}}{t+T_{young}}
\end{equation}
After the merging, the new particle has a mass-loss rate that can be written:
\begin{equation}
\dot{M}_{new}(t)=\frac{c_{new}}{t+T_{new}}
\end{equation}
We just have to determine the parameters $c_{new}$ and $T_{new}$. We impose the 
continuity of the total stellar mass-loss rate at the instant $t$ of the fusion, 
which just reads:
\begin{equation}
\frac{c_{new}}{t+T_{new}} = \frac{c_{young}}{t+T_{young}} + \frac{c_{old}}{t+T_{old}}
\end{equation}
and the conservation of the total stellar 
mass-loss integrated over $20$ Gyr. These two equations determine
the stellar mass-loss 
of a particle after the accretion of some young stars material.
\end{itemize}

Once the stellar mass-loss is computed, we distribute this mass
over the neighboring gas particles.
We use a $1024\times1024$ 
grid to compute the mass transfer in each cell. 
This mass is taken away from the star particles and transfered to the 
gas ones, in proportion to their mass, and the transfer of linear momentum that is associated with this mass redistribution is computed, just as was done for star formation.

These processes are not taken
into account at the same frequency as the particle advance, and
gravity potential computations.
The time steps for stellar mass-loss and star formation are:
\begin{equation}
\Delta t _{ML}= \Delta t _{SF} = 4 \Delta t _{\phi}
\end{equation}

        \subsection{Gas accretion} \label{accreting}
To represent gas accretion in the two-dimensional code, 
some assumptions are necessary. 
We assume that the mean speed of accreted particles 
is equal to that of the outer disk, considering that slower 
particles have been accreted earlier, and that the accreted gas 
random velocity corresponds to 
the same Toomre parameter as the disk ($Q=1.5$).

Accreted particles are placed just within the disk outer limit. None are put outside 
the disk. So, the disk radius is not directly changed in the process. 
The disk radius however increases indirectly, due 
to the disk response to accretion (see Sect.~3). The disk radius is then computed
as a function of time, in order to continuously place new particles 
just below this radius, and not in the middle of the disk after some Gyr.

The accretion rate, $\dot{M}_{accretion}$, is chosen to make the disk mass double 
in $7$ to $10$ Gyr (see for instance likely accretion rates, derived
by Jiang \& Binney 1999).
The gas is accreted  from large-scale material. However,
in some environments, it is possible that the accreted mass depletes
a significant fraction of the halo mass. Yet, we will note that the 
existence or not of a halo mass depletion does not influence 
the disk evolution. We then generally consider 
that the accreted mass is partly taken away from the halo 
($\dot{M}_{halo}=-\dot{M}_{accret}/3$), excepted in a few runs which are noticed 
in table \ref{param1}.

The consequences of all these assumptions are discussed in Sect.~\ref{discu}.

        \subsection{Three-dimensional code}
	\subsubsection{Disk, gravitation, and star formation}
The 3-D model is quite similar to the 2-D one, as far as the spatial
distributions projected on the plane are concerned. In addition
the mass distribution of the initial disk in the $z$ direction is:
\begin{equation}
\rho(r,z)=\sigma(r) \; \mathrm{sech}^2 \left( \frac{z}{h_0} \right)
\end{equation}
where the disk scale height is $h_0 = 1.5$ kpc, and $\sigma(r)$ is the disk surface density.

The gravitational potential is still obtained by an FFT method. We now use 
a useful grid of $128\times128\times128$. Proceeding Fourier transforms on 
a grid doubled in each dimension, as was done in the 2-D code, is not the 
best method as regards the computational efficiency. Instead we use the 
James (1977) method: appropriate masses are distributed on the grid 
boundaries to compensate for the Fourier periodic images.

Star formation and stellar mass-loss are included in this code with the same schemes as in the 2-D one.

	\subsubsection{Sticky-particles code}
The two-dimensional gas dissipation scheme cannot really be used in
three-dimensional simulations: computing the local mean speed in
each cell implies that there are several gaseous particles in 
each cell. This would compel us either to use a few million 
particles, which means low computational efficiency, or to 
reduce the number of cells, which would make the spatial resolution 
reach a few kpc. Thus, we choose to use the sticky-particles algorithm
(Schwarz, 1981). When two gaseous particles collide, their radial (along the line joining them) and tangential relative velocities are reduced by factors $\beta_r$ and $\beta_t$. We choose $\beta _r=0.65$ and $\beta _t=0.65$ (see Jungwiert \& Palou$\breve{\mathrm{s}}$, 1996).
Let us consider two particles at $\vec{r}_1$ and $\vec{r}_2$, with masses $m_1$ and $m_2$, and velocities $\vec{v}_1$ and $\vec{v}_2$ in the referential of their mass center . We note $M=m_1+m_2$, and:
\begin{equation}\label{stick1}
\vec{e}_r=\frac{\vec{r}_2-\vec{r}_1}{||\vec{r}_2-\vec{r}_1||}
\end{equation}
Before the collision, the radial et tangential relative velocities are respectively:
\begin{equation}
\vec{v}_r=\left((\vec{v}_2-\vec{v}_1)\cdot\vec{e}_r\right)\vec{e}_r
\end{equation}
\begin{equation}
\vec{v}_t=(\vec{v}_2-\vec{v}_1)-\left((\vec{v}_2-\vec{v}_1)\cdot\vec{e}_r\right)\vec{e}_r
\end{equation}
After the collision, they are:
\begin{equation}
\vec{v'}_r=\beta_r \vec{v}_r
\end{equation}
\begin{equation}
\vec{v'}_t=\beta_t \vec{v}_t
\end{equation}
Then the new velocities of the particles in the referential of the mass center are:
\begin{equation}
\vec{v'}_1=\frac{m_2}{M} \left(\beta_t(\vec{v}_1-\vec{v}_2) + (\beta_r-\beta_t)\left((\vec{v}_1-\vec{v}_2)\cdot\vec{e}_r\right)\vec{e}_r\right)
\end{equation}
\begin{equation}\label{stick2}
\vec{v'}_2=\frac{m_1}{M} \left(\beta_t(\vec{v}_2-\vec{v}_1) + (\beta_r-\beta_t)\left((\vec{v}_2-\vec{v}_1)\cdot\vec{e}_r\right)\vec{e}_r\right)
\end{equation}
The energy lost after the collision is then:
\begin{equation}
\Delta E=\left( 1-\beta_r^2\right) \vec{v}_r^2+\left(1-\beta_t^2\right) \vec{v}_t^2
\end{equation}
We will discuss the influence of $\beta_r$ and $\beta_t$ on the results of our simulations in appendix \ref{append}.

	\subsubsection{gas accretion}
There are now many free parameters for gas accretion: the accreted
gas angular momentum direction may be varied, and the accreted gas may
be placed around any place of the disk, while in the 2-D code, it was always
placed around the disk radius. We will precise which accretion
parameters were used in each run when we study the 3-D results, in
Sect.~\ref{3D}.

\section{Two-dimensional effects of gas accretion} \label{sec3}

        \subsection{General observations} \label{gen}
                \subsubsection{Gas accretion and bar evolution}
Depending essentially on bulge to disk and halo to disk mass ratios,
there are two kinds of bar evolution in 2-D simulations. We now
describe them in turn.

\begin{table*}
        \begin{center}
        \begin{tabular}{lccccccc}
        \hline
	\hline
Run  &   $a_{disk}$ &   $a_{halo}$ &   $M_{halo}^{inner}$ &  $M_
{bulge}$ &  $\dot{M}_{accret}$ & $\dot{M_{halo}}$ &initial gas mass rate\\
        \hline
    \\
F1 & 4 kpc & 25 kpc & 2.5 10$^{10}$ M$_{\sun}$& 1.3 10$^{10}$
M$_{\sun}$& 0.6 10$^{10}$ M$_{\sun}$ Gyr$^{-1}$ &
$-\dot{M}_{accret}/2$ & 25 $\%$\\
0F1 & 4 kpc & 25 kpc & 2.5 10$^{10}$ M$_{\sun}$& 1.3 10$^{10}$ M$_{\sun}$& 0 & $-\dot{M}_{accret}/2$& 25 $\%$\\
F2 & 4 kpc & 25 kpc & 4.0 10$^{10}$ M$_{\sun}$& 4.0 10$^{10}$ M$_{\sun}$& 0.6 10$^{10}$ M$_{\sun}$ Gyr$^{-1}$       & $-\dot{M}_{accret}/2$& 25 $\%$\\
0F2 & 4 kpc & 25 kpc & 4.0 10$^{10}$ M$_{\sun}$& 4.0 10$^{10}$ M$_{\sun}$&0       & $-\dot{M}_{accret}/2$& 25 $\%$\\
    \\
A & 4 kpc & 25 kpc &  2.5 10$^{10}$ M$_{\sun}$ & 3.0 10$^{10}$
M$_{\sun}$ &  0.57 10$^{10}$ M$_{\sun}$ Gyr$^{-1}$  &
$-\dot{M}_{accret}/3$ & 20 $\%$\\
0A & 4 kpc & 25 kpc &  2.5 10$^{10}$ M$_{\sun}$ & 3.0 10$^{10}$
M$_{\sun}$ &  0 & $-\dot{M}_{accret}/3$ & 20 $\%$\\
B & 4 kpc & 25 kpc &  3.5 10$^{10}$ M$_{\sun}$ & 1.4 10$^{10}$
M$_{\sun}$ &  0.57 10$^{10}$ M$_{\sun}$ Gyr$^{-1}$  &
$-\dot{M}_{accret}/3$  & 20 $\%$\\
0B & 4 kpc & 25 kpc &  3.5 10$^{10}$ M$_{\sun}$ & 1.4 10$^{10}$
M$_{\sun}$ &  0  & $-\dot{M}_{accret}/3$  & 20 $\%$\\
     \\
A1 & 4 kpc & 25 kpc &  2.0 10$^{10}$ M$_{\sun}$ &  1.7 10$^{10}$ M$_{\sun}$ &  0.5 10$^{10}$ M$_{\sun}$ Gyr$^{-1}$  & $-\dot{M}_{accret}$ & 20 $\%$\\
A1h & 4 kpc & 25 kpc & 2.0  10$^{10}$ M$_{\sun}$ &  1.7 10$^{10}$ M$_{\sun}$ &  0.5 10$^{10}$ M$_{\sun}$ Gyr$^{-1}$  & 0 & 20 $\%$\\
A2 & 4 kpc & 25 kpc &  2.0 10$^{10}$ M$_{\sun}$ &  2.5 10$^{10}$ M$_{\sun}$ &  0.5 10$^{10}$ M$_{\sun}$ Gyr$^{-1}$  & $-\dot{M}_{accret}/3$ & 20 $\%$\\
A3 & 4 kpc & 25 kpc &  2.0 10$^{10}$ M$_{\sun}$ & 3.5 10$^{10}$ M$_{\sun}$ &  0.5 10$^{10}$ M$_{\sun}$ Gyr$^{-1}$  & $-\dot{M}_{accret}/3$ & 20 $\%$\\
A4 & 4 kpc & 25 kpc &  3.0 10$^{10}$ M$_{\sun}$ & 2.5 10$^{10}$ M$_{\sun}$ &  0.5 10$^{10}$ M$_{\sun}$ Gyr$^{-1}$  & $-\dot{M}_{accret}$ & 20 $\%$\\
A4h & 4 kpc & 25 kpc & 3.0  10$^{10}$ M$_{\sun}$ & 2.5 10$^{10}$ M$_{\sun}$ &  0.5 10$^{10}$ M$_{\sun}$ Gyr$^{-1}$  & 0 & 20 $\%$\\
A5 & 4 kpc & 25 kpc &  3.0 10$^{10}$ M$_{\sun}$ & 3.5 10$^{10}$ M$_{\sun}$ &  0.5 10$^{10}$ M$_{\sun}$ Gyr$^{-1}$  & $-\dot{M}_{accret}/3$ & 20 $\%$\\
A6 & 4 kpc & 25 kpc &  4.5 10$^{10}$ M$_{\sun}$ & 2.5 10$^{10}$ M$_{\sun}$ &  0.5 10$^{10}$ M$_{\sun}$ Gyr$^{-1}$  & $-\dot{M}_{accret}/3$ & 20 $\%$\\
A7 & 4 kpc & 25 kpc &  4.5 10$^{10}$ M$_{\sun}$ & 3.5 10$^{10}$ M$_{\sun}$ &  0.5 10$^{10}$ M$_{\sun}$ Gyr$^{-1}$  & $-\dot{M}_{accret}/3$ & 20 $\%$\\
A8 & 4 kpc & 25 kpc &  5.5 10$^{10}$ M$_{\sun}$ & 2.5 10$^{10}$ M$_{\sun}$ &  0.5 10$^{10}$ M$_{\sun}$ Gyr$^{-1}$  & $-\dot{M}_{accret}$ & 20 $\%$\\
A8h & 4 kpc & 25 kpc & 5.5 10$^{10}$ M$_{\sun}$ & 2.5 10$^{10}$ M$_{\sun}$ &  0.5 10$^{10}$ M$_{\sun}$ Gyr$^{-1}$  & 0 & 20 $\%$\\
A9 & 4 kpc & 25 kpc &  5.5 10$^{10}$ M$_{\sun}$ & 3.5 10$^{10}$ M$_{\sun}$ &  0.5 10$^{10}$ M$_{\sun}$ Gyr$^{-1}$  & $-\dot{M}_{accret}/3$ & 20 $\%$\\
     \\
B1 & 4 kpc & 25 kpc &  2.0 10$^{10}$ M$_{\sun}$ &  1.0 10$^{10}$ M$_{\sun}$ &  0.5 10$^{10}$ M$_{\sun}$ Gyr$^{-1}$  & $-\dot{M}_{accret}/3$ & 20 $\%$\\
B2 & 4 kpc & 25 kpc &  3.0 10$^{10}$ M$_{\sun}$ &  1.0 10$^{10}$ M$_{\sun}$ &  0.5 10$^{10}$ M$_{\sun}$ Gyr$^{-1}$  & $-\dot{M}_{accret}/3$ & 20 $\%$\\
B3 & 4 kpc & 25 kpc &  4.0 10$^{10}$ M$_{\sun}$ &  1.7 10$^{10}$ M$_{\sun}$ &  0.5 10$^{10}$ M$_{\sun}$ Gyr$^{-1}$  & $-\dot{M}_{accret}/3$ & 20 $\%$\\
B4 & 4 kpc & 25 kpc &  4.5 10$^{10}$ M$_{\sun}$ &  1.0 10$^{10}$ M$_{\sun}$ &  0.5 10$^{10}$ M$_{\sun}$ Gyr$^{-1}$  & $-\dot{M}_{accret}/3$ & 20 $\%$\\
B5 & 4 kpc & 25 kpc &  4.5 10$^{10}$ M$_{\sun}$ &  1.7 10$^{10}$ M$_{\sun}$ &  0.5 10$^{10}$ M$_{\sun}$ Gyr$^{-1}$  & $-\dot{M}_{accret}$ & 20 $\%$\\
B5h & 4 kpc & 25 kpc &  4.5 10$^{10}$ M$_{\sun}$ &  1.7 10$^{10}$ M$_{\sun}$ &  0.5 10$^{10}$ M$_{\sun}$ Gyr$^{-1}$  & 0 & 20 $\%$\\
B6 & 4 kpc & 25 kpc &  5.5 10$^{10}$ M$_{\sun}$ &  1.0 10$^{10}$ M$_{\sun}$ &  0.5 10$^{10}$ M$_{\sun}$ Gyr$^{-1}$  & $-\dot{M}_{accret}/3$ & 20 $\%$\\
B7 & 4 kpc & 25 kpc &  5.5 10$^{10}$ M$_{\sun}$ &  1.7 10$^{10}$ M$_{\sun}$ &  0.5 10$^{10}$ M$_{\sun}$ Gyr$^{-1}$  & $-\dot{M}_{accret}/3$ & 20 $\%$\\
    \\
R1a & 4 kpc & 25 kpc &  2.5 10$^{10}$ M$_{\sun}$ &  1.5 10$^{10}$ M$_{\sun}$ &  0.6 10$^{10}$ M$_{\sun}$ Gyr$^{-1}$  & $-\dot{M}_{accret}/3$ & 20 $\%$\\
R1b & 4 kpc & 25 kpc &  4.0 10$^{10}$ M$_{\sun}$ &  2.5 10$^{10}$ M$_{\sun}$ &  0.6 10$^{10}$ M$_{\sun}$ Gyr$^{-1}$  & $-\dot{M}_{accret}/3$ & 20 $\%$\\
R1c & 4 kpc & 25 kpc &  5.5 10$^{10}$ M$_{\sun}$ &  3.5 10$^{10}$ M$_{\sun}$ &  0.6 10$^{10}$ M$_{\sun}$ Gyr$^{-1}$  & $-\dot{M}_{accret}/3$ & 20 $\%$\\
R2a & 5 kpc & 20 kpc &  2.5 10$^{10}$ M$_{\sun}$ &  1.5 10$^{10}$ M$_{\sun}$ &  0.6 10$^{10}$ M$_{\sun}$ Gyr$^{-1}$  & $-\dot{M}_{accret}/3$ & 20 $\%$\\
R2b  & 5 kpc & 20 kpc &  4.0 10$^{10}$ M$_{\sun}$ &  2.5 10$^{10}$ M$_{\sun}$ &  0.6 10$^{10}$ M$_{\sun}$ Gyr$^{-1}$  & $-\dot{M}_{accret}/3$ & 20 $\%$\\
R2c & 5 kpc & 20 kpc &  5.5 10$^{10}$ M$_{\sun}$ &  3.5 10$^{10}$ M$_{\sun}$ &  0.6 10$^{10}$ M$_{\sun}$ Gyr$^{-1}$  & $-\dot{M}_{accret}/3$ & 20 $\%$\\
R3a & 6 kpc & 15 kpc &  2.5 10$^{10}$ M$_{\sun}$ &  1.5 10$^{10}$ M$_{\sun}$ &  0.6 10$^{10}$ M$_{\sun}$ Gyr$^{-1}$  & $-\dot{M}_{accret}/3$ & 20 $\%$\\
R3b  & 6 kpc & 15 kpc &  4.0 10$^{10}$ M$_{\sun}$ &  2.5 10$^{10}$ M$_{\sun}$ &  0.6 10$^{10}$ M$_{\sun}$ Gyr$^{-1}$  & $-\dot{M}_{accret}/3$ & 20 $\%$\\
R3c & 6 kpc & 15 kpc &  5.5 10$^{10}$ M$_{\sun}$ &  3.5 10$^{10}$ M$_{\sun}$ &  0.6 10$^{10}$ M$_{\sun}$ Gyr$^{-1}$  & $-\dot{M}_{accret}/3$ & 20 $\%$\\
    \\
G1 & 4 kpc & 25 kpc &  2.5 10$^{10}$ M$_{\sun}$ &  2.0 10$^{10}$ M$_{\sun}$ &  0.5 10$^{10}$ M$_{\sun}$ Gyr$^{-1}$  & $-\dot{M}_{accret}/3$ & 5 $\%$\\
G2 & 4 kpc & 25 kpc &  2.5 10$^{10}$ M$_{\sun}$ &  2.0 10$^{10}$ M$_{\sun}$ &  0.5 10$^{10}$ M$_{\sun}$ Gyr$^{-1}$  & $-\dot{M}_{accret}/3$ & 15 $\%$\\
G3 & 4 kpc & 25 kpc &  5.5 10$^{10}$ M$_{\sun}$ &  2.0 10$^{10}$ M$_{\sun}$ &  0.5 10$^{10}$ M$_{\sun}$ Gyr$^{-1}$  & $-\dot{M}_{accret}/3$ & 25 $\%$\\
G4 & 4 kpc & 25 kpc &  2.5 10$^{10}$ M$_{\sun}$ &  2.0 10$^{10}$ M$_{\sun}$ &  0.5 10$^{10}$ M$_{\sun}$ Gyr$^{-1}$  & $-\dot{M}_{accret}/3$ & 40 $\%$\\
        \hline
        \end{tabular}
        \end{center}
        \caption{Run parameters. The initial disk mass is always
          M$_{disk}^{initial}=4\cdot 10^{10}\;\mathrm{M}_{\sun}$. The
          runs F1 and F2 are introduced in Sect.~\ref{gen}. The
          runs A and B are studied in Sect.~\ref{det}. Runs A1 to
          A9 and B1 to B7 are used to determine the dependence of bar
          evolution on parameters ; in terms of bar evolution, these
          runs are rather similar either to run A or to run B. Runs
          R1a to R3c allow us to study the link with the de
          Vaucouleurs classification (see Sect.~\ref{arms}).}
        \label{param1}
\end{table*}

                        \paragraph{Bar maintenance}
\begin{figure*}
%        \includegraphics[width=4.25cm]{d0_0.png}
%        \includegraphics[width=4.25cm]{d0_2500.png}
%        \includegraphics[width=4.25cm]{d0_5000.png}
%        \includegraphics[width=4.25cm]{d0_7500.png}\\
%        \includegraphics[width=4.25cm]{d0_10000.png}
%        \includegraphics[width=4.25cm]{d0_12500.png}
%        \includegraphics[width=4.25cm]{d0_15000.png}
%        \includegraphics[width=4.25cm]{d0_17500.png}\\
%        \medskip\\
%        \includegraphics[width=4.25cm]{d1_0.png}
%        \includegraphics[width=4.25cm]{d1_2500.png}
%        \includegraphics[width=4.25cm]{d1_5000.png}
%        \includegraphics[width=4.25cm]{d1_7500.png}\\
%        \includegraphics[width=4.25cm]{d1_10000.png}
%        \includegraphics[width=4.25cm]{d1_12500.png}
%        \includegraphics[width=4.25cm]{d1_15000.png}
%        \includegraphics[width=4.25cm]{d1_17500.png}
\caption{Runs F1 and 0F1: disk evolution with accretion and bar maintenance (F1, top), and without accretion (0F1, bottom).}
\label{disk1}
\end{figure*}

Let us first examine the model of run F1, the parameters of which are given in table \ref{param1}. 
The disk evolution is plotted in Fig.~\ref{disk1}.

Without gas accretion, this disk develops a bar, with a short-lifetime. The bar 
is very strong around $1$ and $2$ Gyr. The central condensation rises very fast, 
and destroys the bar in less than $7$ Gyr. The introduction of gas accretion makes 
the bar more long-lived. The disk has a bar wave during more than $15$ Gyr. 
The bar destruction process still occurs, but is much slower. Therefore, one 
of the main effects of gas accretion is to maintain bars in such galactic disks.

Moreover, the maintained bar can still be strong enough
to correspond to a morphological SB-type, after more than $10$ Gyr. 
A SB-type disk or strongly barred disk is a disk in which the bar is very long, with 
two arms connected to the bar. In a SAB-type disk, the bar is less long 
and has rather a lens aspect ; the spiral structure is rather disconnected from 
the bar.

A strong bar is able to influence the spiral pattern outside of its
corotation, so that the spiral and the bar have the same pattern speed.
In the run F1, at $15$ Gyr, the spiral pattern appears always in phase with  
the bar, as in SB-type galaxies.
Even if the bar is stronger during the first Gyrs, the disk may become 
more weakly barred, as an SAB-type, and return 
to SB-type with an older bar. The frequencies of SB and SAB-type disks are 
studied in Sect.~\ref{arms}.

Runs B and B1 to B7 also show such a bar maintenance. The bar lifetime for each 
of them is described in table \ref{modB}

\begin{table}
        \begin{center}
        \begin{tabular}{lc}
       \hline
       \hline
Run  &  Bar lifetime\\
        \hline
B1 & 13.5  \\
B2 & 15  \\
B3 & 17  \\
B4 & 12.5  \\
B5 & 19  \\
B5h& 18.5  \\
B6 & 13  \\
B7 & 15.5  \\
        \hline
        \end{tabular}
        \end{center}
        \caption{Bar lifetime for bar maintaining galaxies.}
        \label{modB}
\end{table}

                \paragraph{Bar destruction and re-formation}

\begin{figure*}
%        \includegraphics[width=4.25cm]{d2_0.png}
%        \includegraphics[width=4.25cm]{d2_2000.png}
%        \includegraphics[width=4.25cm]{d2_5000.png}
%        \includegraphics[width=4.25cm]{d2_8000.png}\\
%        \includegraphics[width=4.25cm]{d2_11000.png}
%        \includegraphics[width=4.25cm]{d2_14000.png}
%        \includegraphics[width=4.25cm]{d2_17000.png}
%        \includegraphics[width=4.25cm]{d2_20000.png}\\
%        \medskip\\
%        \includegraphics[width=4.25cm]{d3_0.png}
%        \includegraphics[width=4.25cm]{d3_2000.png}
%        \includegraphics[width=4.25cm]{d3_5000.png}
%        \includegraphics[width=4.25cm]{d3_8000.png}\\
%        \includegraphics[width=4.25cm]{d3_11000.png}
%        \includegraphics[width=4.25cm]{d3_14000.png}
%        \includegraphics[width=4.25cm]{d3_17000.png}
%        \includegraphics[width=4.25cm]{d3_20000.png}\\
\caption{Runs F2 and 0F2: disk evolution with accretion and bar destruction 
and re-formation (F2, top), and without accretion (0F2, bottom). Here, the bar 
is totally destroyed before its re-formation. It can also be much weakened 
without disappearing (see Fig.~\ref{strengthAB} for such a situation in model 
A). Two bar episodes can be seen here in 20 Gyrs ; three or four bar episodes 
have been followed as well in similar runs (A1-A9).} 
\label{disk2}
\end{figure*}

The second situation, obtained for instance with parameters of run F2 
from table \ref{param1}, 
corresponds to bar destruction and re-formation. The disk evolution of this run is shown in 
Fig.~\ref{disk2}.

Now, the isolated disk has a strong bar during less than $2$ Gyr. This
bar destroys itself very fast. Then, a small lens-shaped bar
remains in the inner disk, which testifies a previous strong bar existence. Even if
the real bar is a short life-time one, the lens is observed over more
than 10 Gyr.

Gas accretion allows the bar to reappear. After being totally unbarred
-even the lens is now destructed-, the disks undergoes a new bar episode. 
A cycle of bar destruction 
and re-formation takes places in such disks. We generally obtain $2$ or $3$ cycles 
in $20$ Gyr, sometimes $4$.

If the strength of the bar tends to decrease with time, because of the mass concentration,
and the disk heating (larger velocity dispersion), there still can exist
quite strong bars (SB-types) after $10$ Gyr. Due to gas accretion, and 
consequent disk cooling, 
the second or the third bars can be strong, almost as strong as the first bar. 

In runs A and A1 to A9, we also follow such bar destructions and
re-formations. The number of bars followed over 20 Gyrs for these runs 
is given in table \ref{modA}.

\begin{table}
        \begin{center}
        \begin{tabular}{lc}
       \hline
       \hline
Run  &  Number of bars in 20 Gyrs \\
        \hline
A1 & 4 \\
A1h& 4  \\
A2 & 2  \\
A3 & 2  \\
A4 & 3  \\
A4h& 3  \\
A5 & 2  \\
A6 & 3  \\
A7 & 2  \\
A8 & 3  \\
A8h& 3  \\
A9 & 2  \\
        \hline
        \end{tabular}
        \end{center}
        \caption{Number of successive bars followed over 20 Gyrs in bar 
re-forming galaxies.}
        \label{modA}
\end{table}
                \paragraph{Dependence of bar evolution on parameters}
A galactic disk may either keep a single bar or get several successive bars. By 
simulating several disks, we study the dependence on physical parameters. We find 
that the two determining parameters are the bulge to disk and halo to disk masses 
ratios. Light bulge galaxies maintain their bar, while bars are destructed and 
regenerated in heavy bulge galaxies. The halo mass has also an effect on the disk 
evolution. The bar evolution of a galactic disk, depending on bulge and halo to 
disk masses ratios, is shown on Fig.~\ref{params}.

The bulge mass also influences the speed of the destruction and re-formation cycle, 
in bar re-forming galaxies. It influences the bar lifetime in bar maintaining 
galaxies. Both influences are explained on Fig.~\ref{params}.

The initial gas to stellar disk mass ratio has also been varied (runs G1-G4). The
higher it is, the stronger the density waves are during the 1 or 2
first Gyr. Yet, this ratio does not really influence the long-term
evolution of the disk: After a few Gyr, the gas to stellar mass ratio
depends more on the accretion rate rather than on the initial gas to 
stellar mass ratio.

\begin{figure}
\resizebox{8cm}{!}{\includegraphics{param.eps}}

 {\includegraphics[width=.26cm]{symb1.eps}} Runs B1-B7: Bar maintenance.

[bar lifetime in Gyr]
\medskip

 {\includegraphics[width=.26cm]{symb2.eps}} Runs A1-A9: Bar destruction and 
re-formation.

(number of bars over 20 Gyr)

        \caption{Disk evolution depending on bulge to disk and halo to disk masses 
ratio. $M_{disk}$ is the initial disk mass, and $M_{halo}^{inner}$ is one tenth 
$M_{halo}$.}
        \label{params}
\end{figure}

                \subsubsection{Other consequences of accretion}

The main other consequences of accretion are:
\begin{itemize}
\item the spiral structure is rejuvenated by accretion (see also Sect.~\ref{arms}). 
Arms have roughly the same morphology around $15$ Gyr as around $3$ Gyr, 
whereas in non accreting disks, they become weaker, since the gas 
density then decreases in the disk while the gas accumulates 
in an inner ring and/or an outer ring.
\item the disk radius increases; accreted gas, even if always placed 
inside the disk border, produces larger disks.
\item there is sometimes a gaseous outer ring, which is stronger than without accretion. 
This, as shown below, is due to the accreted gas repelled outside the corotation because 
of bar torques.
\item the bar pattern speed raises; this will be shown and 
interpreted in Sect.~\ref{det}.
\end{itemize}
The three last phenomena are now studied and explained.

        \subsection{Detailed study of two galactic disks} \label{det}

\begin{figure*}
        \centering
        \includegraphics[width=8.5cm]{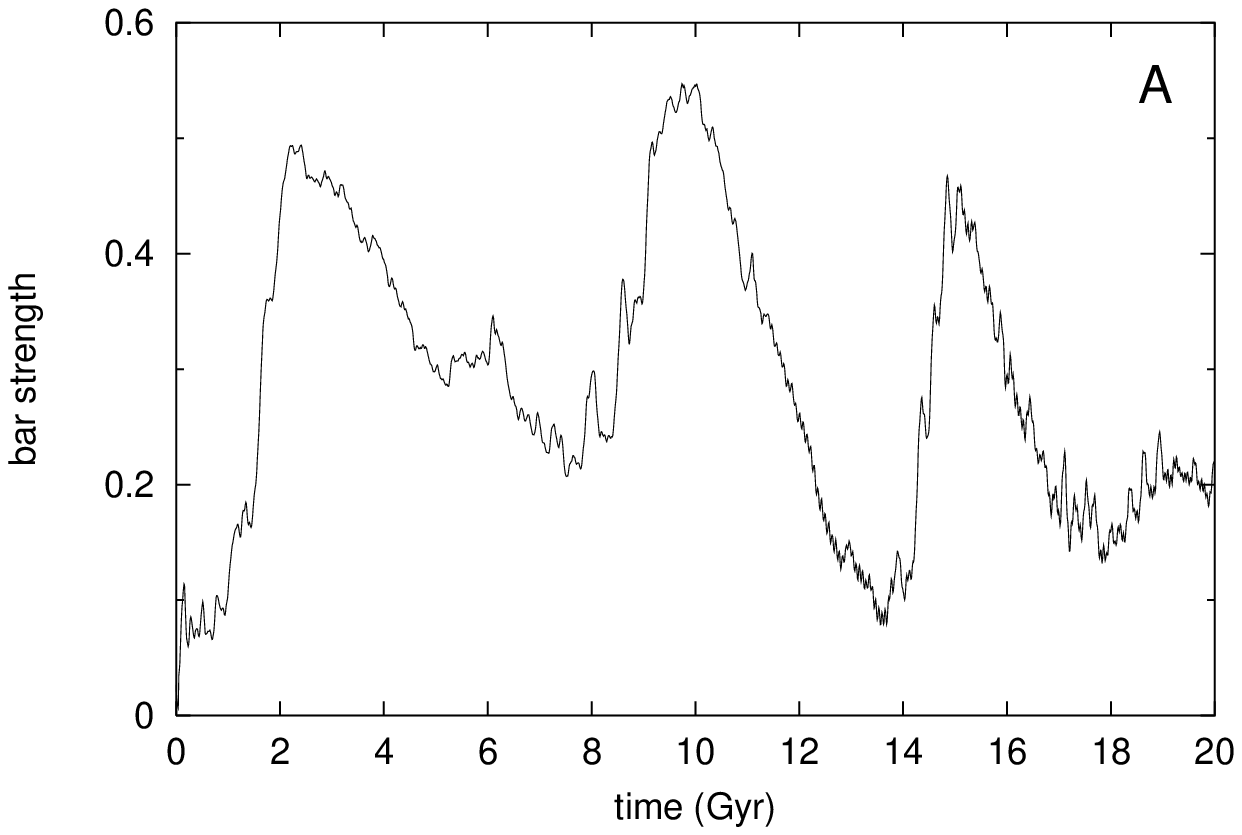}
        \includegraphics[width=8.5cm]{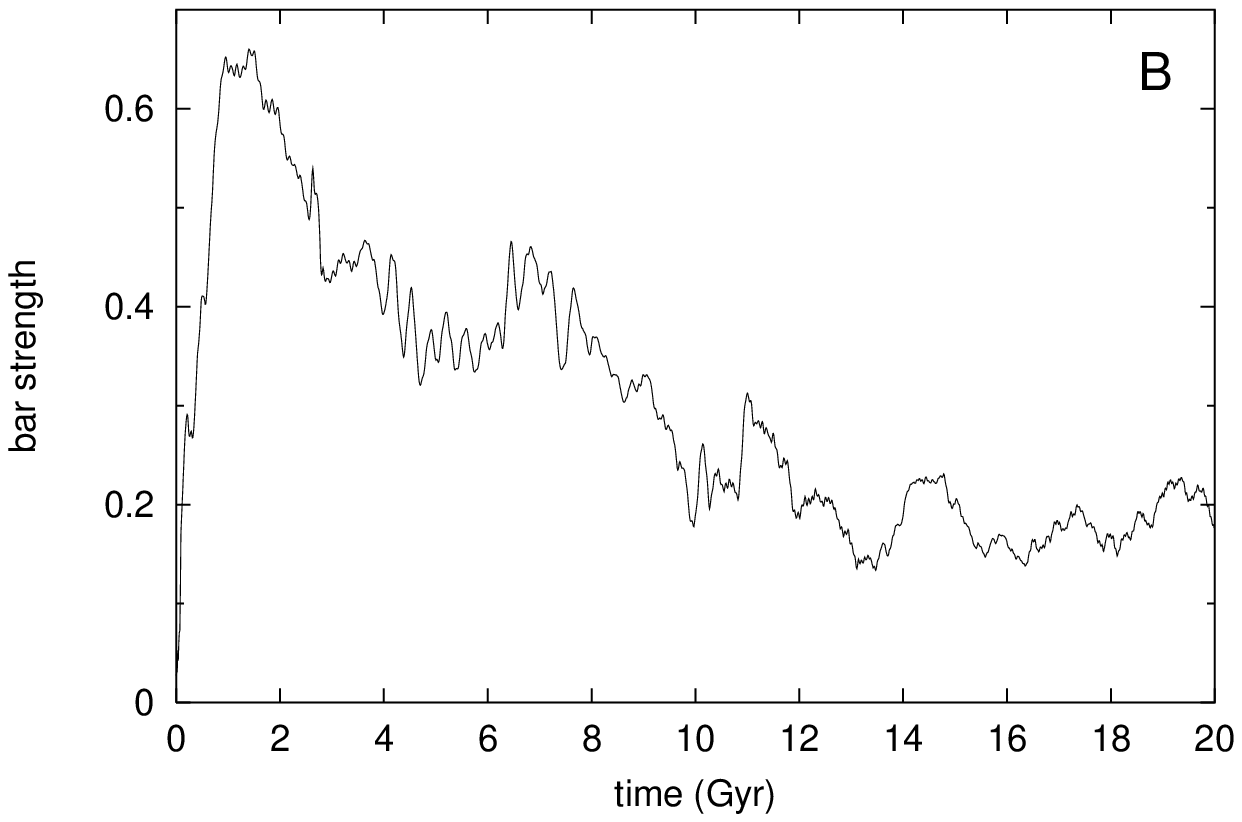}
        \caption{Bar strength evolution in models A (left) and B (right).}
        \label{strengthAB}
\end{figure*}

Let us now present two runs that have been studied in detail, runs A and B in 
table \ref{param1}.

Model A corresponds to a disk which undergoes 
three successive bars, whereas model B corresponds 
to a single bar that persists over a Hubble time. The bar strength evolutions for both 
models are shown in Fig.~\ref{strengthAB}.
The bar strength is defined as the ratio of the tangential 
force of the $m=2$ Fourier component, normalized to the radial force.
If the potential  is decomposed as
\begin{equation}
\Phi(r,\theta) = \Phi_0(r) + \sum_m \Phi_m(r) \cos (m \theta - \phi_m)
\end{equation}
the bar strength is the maximum over radius of 
\begin{equation}
S_2 = 2 \Phi_2 / r | F_r |
\end{equation}
where $ F_r = -\partial \Phi_0 /  \partial$r.

The cause of these two different scenarii in models A and B, the evolution of the bar 
pattern speed, and the increase in the disk radius, are now developed and 
studied.

                \subsubsection{Evolution of the accreted gas distribution}

\begin{figure*}
        \centering
        \caption{Evolution of the radial distribution of accreted during the first 
$1.5$ Gyr gas. Model A is on the left, model B is on the right.}
        \label{distrib}
\end{figure*}

\begin{figure*}
        \centering
        \includegraphics[width=5.7cm]{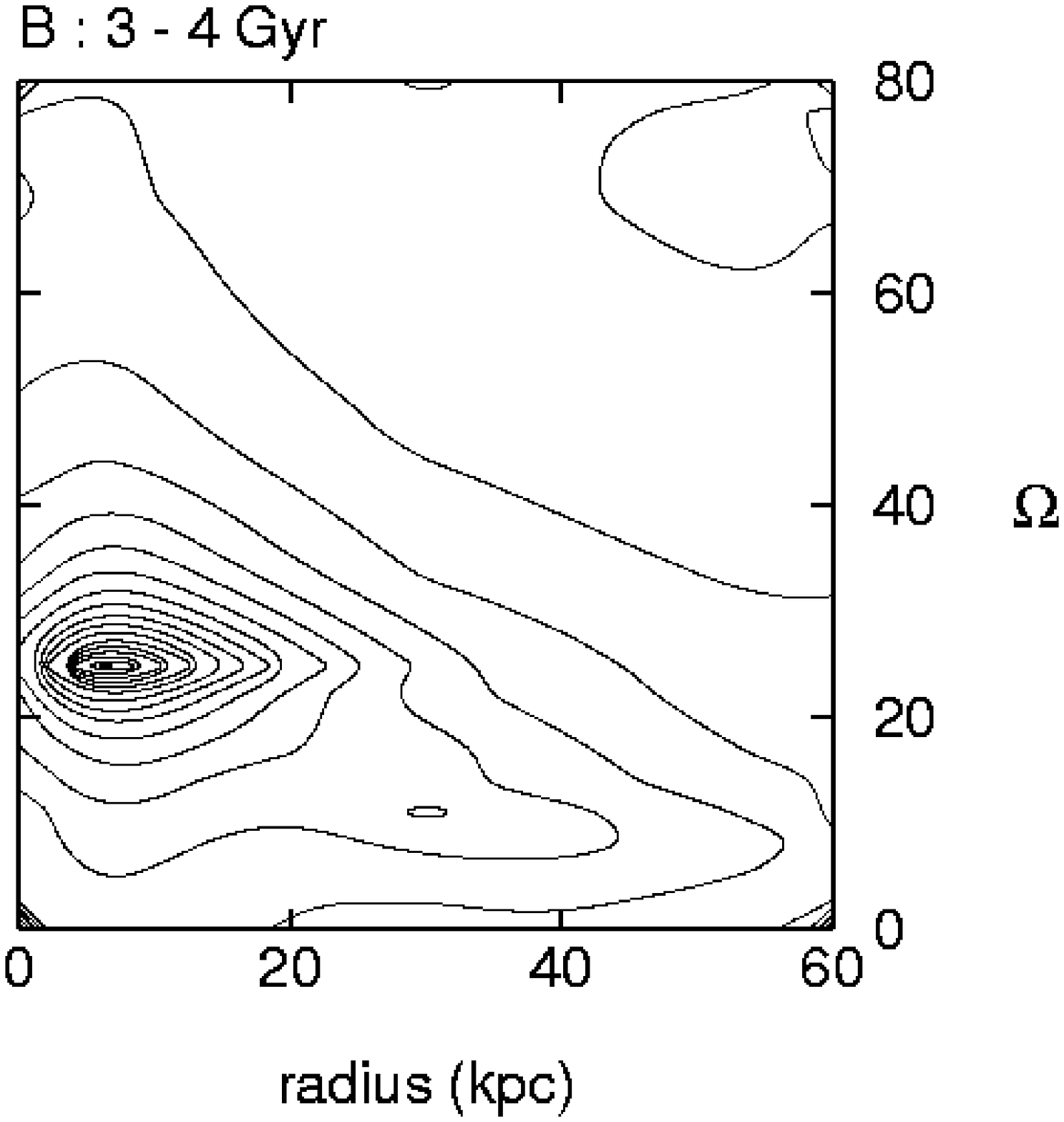}
        \includegraphics[width=5.7cm]{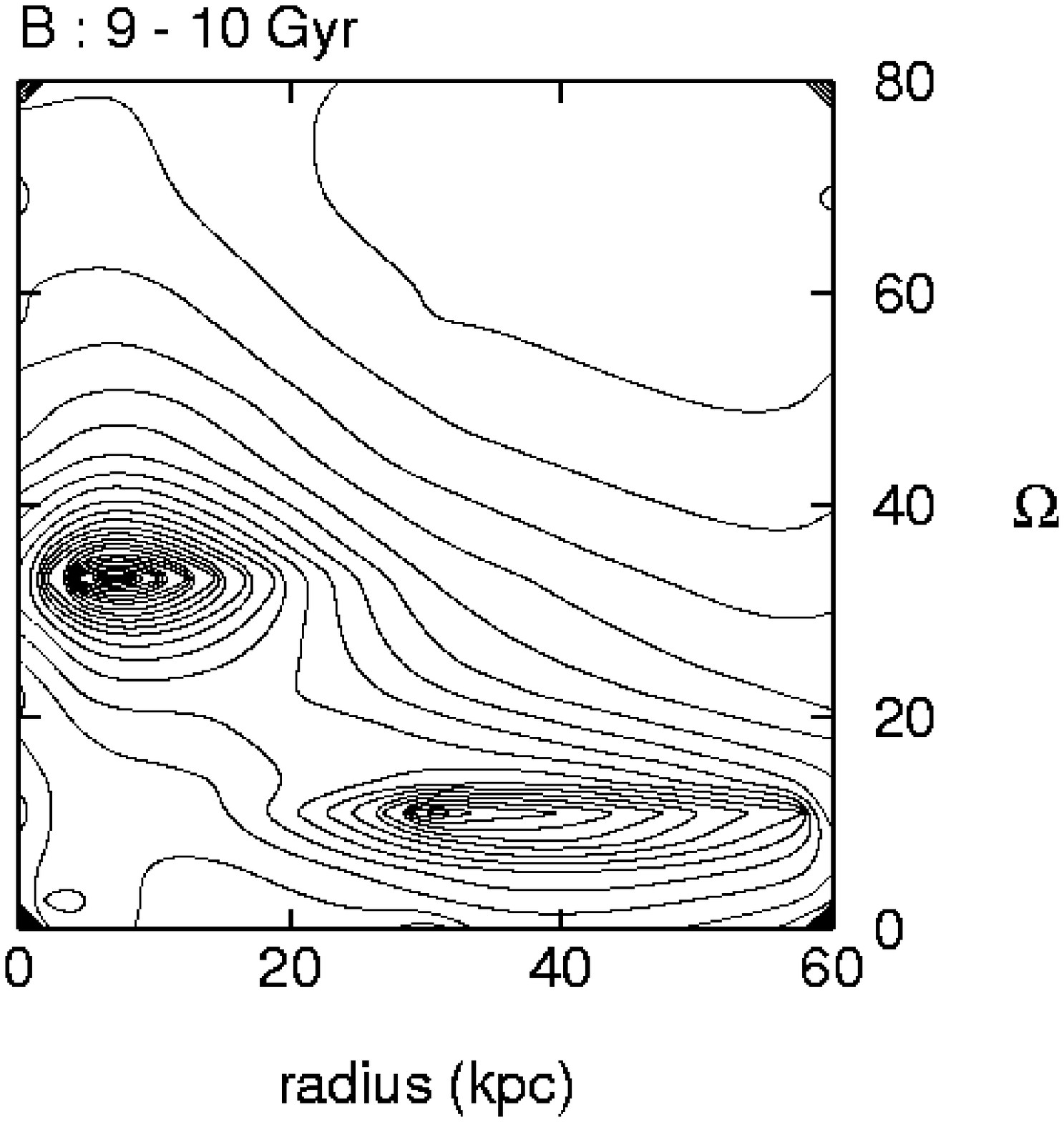}
        \includegraphics[width=5.7cm]{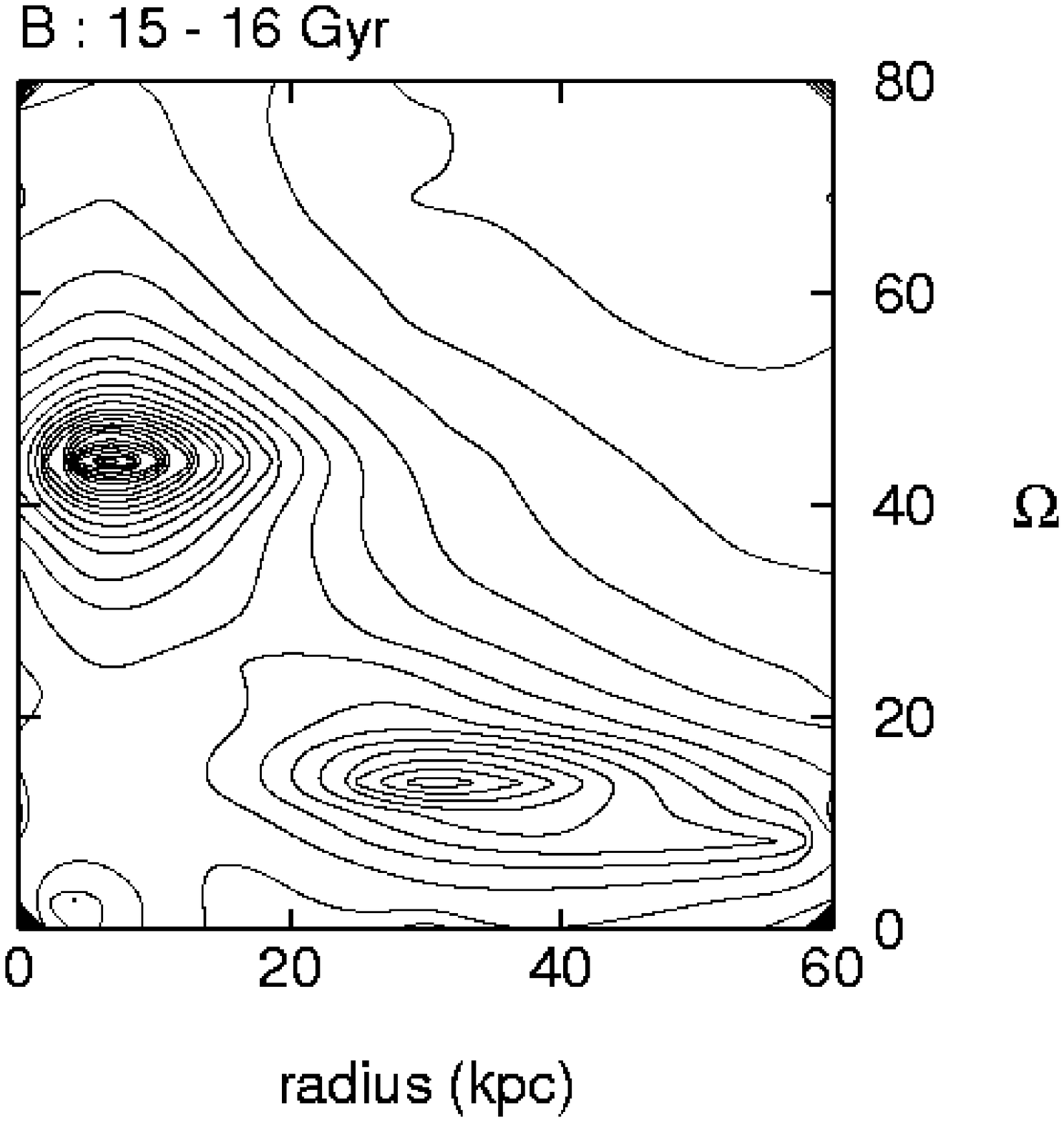}
        \caption{Second harmonic of the potential, as a function of radius and pulsation, 
for model B. The main maximum (at small radii and high frequencies) corresponds to the bar, 
the secondary one (at large radii and low frequencies) to spiral waves. Similar analysis are done for model A.}
        \label{bav}
\end{figure*}

The accreted gas is first located in the outer disk. Yet, non axisymmetric gravity 
fields and dissipation make its distribution evolve. The accreted gas distribution 
and its evolution are shown in Fig.~\ref{distrib}. Only particles that are accreted 
during the first $1.5$ Gyr are taken into account in this figure, in order to have a 
reasonable time axis, as the accretion period of such particles is rather short in 
comparison with the galaxy evolution time. If we consider the particles that are 
accreted between any epoch $t$ and $t+1.5$ Gyr, the results are very similar to 
those plotted in Fig.~\ref{distrib}.

For both models A and B, we see that the accreted gas density is never high in the region
between $4$ and $7$ kpc, which corresponds to the bar corotation in both 
models. Tangential forces exerted by the bar prevent gas from staying in this 
part of the disk. Then, accreted particles may either stay in the outer disk, or 
fall towards the disk center, if some dissipation make them go there. Actually, we see on 
Fig.~\ref{distrib} that accreted gas accumulates around the galactic center after 
some Gyr. To go there, it must overcome bar positive torques. The only mechanism
responsible is the gas dissipation, which becomes higher and higher as gas 
density outside the corotation increases. By spreading out in the disk, some gas 
reaches the corotation. Then, bar negative torques drive it quickly into the 
disk center.

Yet, there are some differences between models A and B. The gas repelled 
outside the corotation seems to be more abundant for model A. The particles get 
over the corotation later, after $8$ Gyr instead of $6$ or $7$ Gyr for model B. This interval 
is rather small, but it is significant that for model B, there is still a strong 
bar when the gas falls inside the corotation, whereas in model A, the corotation 
crossing occurs only when the bar strength is minimum. Moreover, the gas infall 
inside the corotation is rather regular and takes some Gyrs for model B ; for model 
A this infall is faster and happens at once, between the two first bar episodes.

This difference concerning the accreted gas repartition allows us to account for the 
two encountered situations. Moreover, it will explain that, in the next subsections, 
different phenomena occur in a continuous way for model B, and only between bar 
episodes for model A.

                \subsubsection{Explanation of bar destruction and re-formation or maintenance}
The common point between models A and B is that gas accretion favours bars. The 
reason for that is the increase in the disk mass. In non accreting disks, the central 
condensation mass raises, which stabilizes the disk against barred
waves. Accretion 
makes the disk mass increase and become more 
self-gravitating, while making the central (inner disk and bulge) become 
relatively lower, which may either 
prevent the bar from being destroyed or re-form a bar.

The difference takes place in the accreted gas repartition. For model B, this gas falls 
inside the corotation continuously. So, the region of the possible bar (which is $r<4$ 
to $6$ kpc) increases in weight regularly. The bar is then favored regularly, 
and is maintained a long time.

For model A, the effects favourable to bars occur only between the bar episodes. During 
such a bar episode, the accreted gas stays in the outer disk. Even if the total disk mass 
rises, only the outer regions increase in mass: this cannot maintain the bar nor create 
a new one. Between two bar episodes, the region of the possible bar gains mass. 
The possible barred region to central condensation mass ratio becomes lower, and a new 
bar wave appears.

Let us consider now the bar 
pattern speed evolution, and how it is related to the accreted gas 
distribution.

                \subsubsection{Influence of accretion on bar pattern speed}

\begin{figure}
        \centering
        \includegraphics[width=8.5cm]{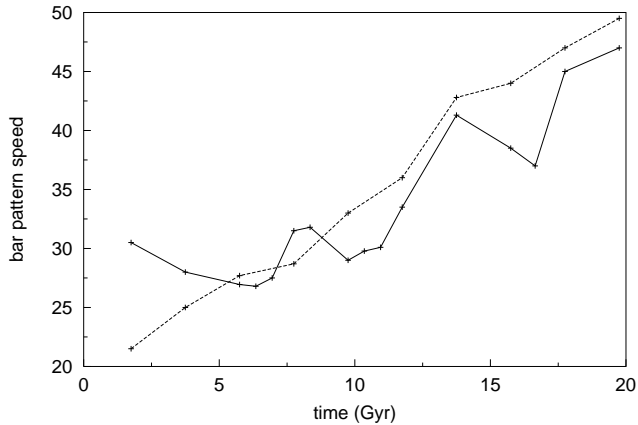}
        \caption{Bar pattern speed as a function of time, for model A (full line) and 
model B (dashed line).}
        \label{bav2}
\end{figure}

The bar pattern speed, $\Omega_b$, has been measured via temporal Fourier 
transform of the $m=2$ harmonic of the potential. Fig.~\ref{bav} shows the 
intensity of this harmonic as a function of radius and frequency $\Omega$, around 
$3.5$, $9.5$  and $15.5$ Gyr. Using such Fourier analysis, we are able to 
summarize $\Omega_b$ as a function of time, as shown in Fig.~\ref{bav2}.

This reveals a significant increase in the bar pattern speed, 
which may double over a Hubble time. The difference between 
the two A and B cases is that this growth is either smooth or irregular. This 
suggests that the bar acceleration is closely linked to gas accretion, 
since it has the same evolution as the accreted gas distribution in both models A and B.

The implications of such a pattern speed increase are developed below. 
Let us first analyze the reasons for this phenomenon.

                \subsubsection{Physical cause of the bar acceleration}
                        \paragraph{Central condensation growth}

A possible cause of the growth in the bar pattern speed is the inner mass accumulation. 
For every particles, this would make $\Omega$ and $\kappa$ rise. If we write 
$M(r)$ the total (disk, bulge and halo) mass that is inside the $r$ radius, all 
frequencies for particles around $r$ are to increase proportionally to $\sqrt{M(r)}$. 
We will study the evolution of $M_{inner}=M(2\;\mathrm{kpc})$, for $2$ kpc is 
generally about the mean radius of the bar components (whereas the maximum of 
the potential harmonics are around the bar extremity).

\begin{figure}
        \centering
        \includegraphics[width=8.15cm]{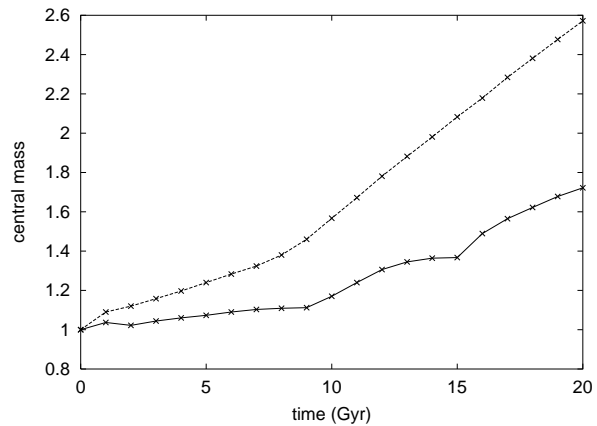}
        \caption{Evolution of the disk inner mass, $M_{disk}(2\mathrm{kpc})$, normalized to its initial value, for model A (full line) and model B (dashed line). Bulge and halo inner masses are to be added to it to give $M_{inner}$.}
        \label{Mr}
\end{figure}

We obtain the expected increase in $M_{inner}$ (see Fig.~\ref{Mr}). The aspect 
of this growth, regular or not, proves that there is a link between it and 
bars succession. For model A, each bar episode makes $\dot{M}_{inner}$ raise: between 
two bars, much accreted mass comes inside the corotation, then the next bar 
exerts tangential forces that make this mass fall into the central condensation. 
For model B, $\dot{M}_{inner}$ is rather constant: accreted gas crosses the 
corotation in a permanent way, and supplies the central condensation during the 
whole galactic evolution. Let us note that $\dot{M}_{inner}$ is considered as constant 
(for model B), while it is actually smaller during the 6 or 7 first Gyrs. Yet, this is explained by 
the absence of particles accreted before $t=0$: most of the actually accreted 
particles only come into the inner disk after $6-7$ Gyr, and only few of them go 
into the central condensation earlier. So, this change in $\dot{M}_{inner}$ is 
linked to the beginning of accretion at $t=0$, it has no physical meaning, and 
what is significant is that $\dot{M}_{inner}$ is constant for $t>7$ Gyr.

Accretion is responsible for an increase in $M_{inner}$. But 
this increase is not sufficient to explain the whole bar acceleration. For 
instance, the growth rates of $\sqrt{M_{inner}}$ and of $\Omega_b$ for model A, 
between several instants, are given in table \ref{Mr_Wb}. The increase 
rate of $\sqrt{M_{inner}}$ is not sufficient, and is only
partly responsible for the increase in $\Omega_b$: the growth rate 
$\Omega_b$ is twice bigger. The bar pattern speed is also influenced by
another phenomenon.

\begin{table}
        \begin{center}
        \begin{tabular}{ccc}
        \hline
	\hline
         growth factor between the first bar and...
                  & $\sqrt{M_{inner}}$ & $\Omega_b$ \\
        \hline
        ...the second bar & $1.06$ & $1.11$ \\
        \hline
        ...the third bar & $1.20$ & $1.37$ \\
        \hline
        \end{tabular}
        \end{center}
        \caption{Inner mass and bar pattern speed growth rates (model A).}
        \label{Mr_Wb}
\end{table}

                        \paragraph{Accretion and angular momentum supply}
\begin{figure}
\resizebox{8cm}{!}{\includegraphics[width=8.15cm]{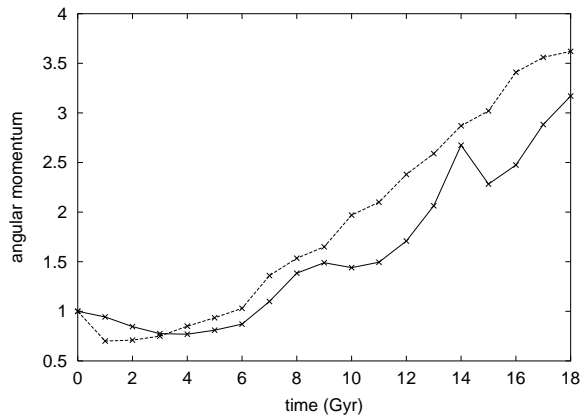}}
        \caption{Inner ($r < 2$ kpc) stellar and gaseous particles angular momentum per unit mass, normalized to its value at $t=0$, for model A (full line) and model B (dashed line).}
        \label{mom}
\end{figure}

In non accreting galactic disks, non axisymmetric fields provoke angular momentum 
transfers from bar to gas outside the corotation, and from gas inside the corotation 
to bar. In model A, the gas density is high outside the corotation when there is 
a strong bar, then the bar is expected to give some momentum to outer particles. 
When the bar strength is lower, much gas arrives in the inner disk, and may bring 
there its angular momentum. This is observed on Fig.~\ref{mom}. For model B, the 
accreted gas density evolves more regularly, and the angular momentum 
per unit mass raises steadily.

Gas accretion has therefore the main
implication to raise the inner angular momentum per unit mass. If $\Omega$ is higher for a 
particle, its orbit's precession rate will also be faster, and it is likely that the
bar instability it will participate in, will have a higher pattern speed. 
The angular momentum supply makes the bar pattern speed raise.

\begin{figure*}
        \caption{Inner disk gaseous and stellar density for model A, seen at $3$, $10$ and $15$ Gyr. The unit of length is $300$ pc.}
        \label{forme}
\end{figure*}

Moreover, gaining angular momentum should make the bar become more circular, because 
faster particles have less eccentric orbits. This is observed on Fig.~\ref{forme} 
for model A, and this phenomenon is similar for model B.
\bigskip

The main implications of gas accretion are therefore an
increase of the central mass and of the angular momentum supply, and a 
higher bar pattern speed. Moreover, the bar becomes rounder as it 
gains angular momentum.

                \subsubsection{Evolution of the disk radius}
 
\begin{figure}
        \centering
        \includegraphics[width=8.15cm]{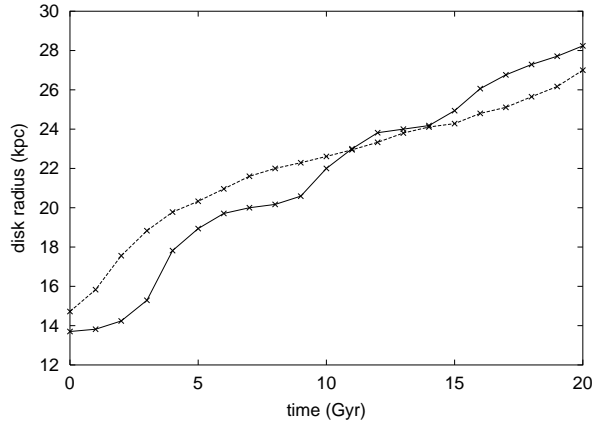}
        \caption{Disk radius as a  function of time, for models A (full line) and B (dashed line).}
        \label{radius}
\end{figure}

The last point to discuss about models A and B is the disk 
radius augmentation. This is plotted on Fig.~\ref{radius}. Here, the disk radius is defined 
as the radius containing  90 percents of the disk mass.

The stronger the bar, the faster the radius grows. For both models, the 
OLR is outside the disk. Yet, bar torques may not explain totally the radius growth, for 
at least two reasons:
\begin{itemize}
\item even if they could enlarge outer particles orbits, they would have the same 
consequences without any gas accretion, which is not observed.
\item when models are studied in which the OLR is inside the disk, the bar torques could 
only make the disk radius decrease. Yet, we still observe a radius increase.
\end{itemize}
The likely explanation is rather that the stronger the bar, the more accreted 
gas is accumulated in the outer disk: this gas spreads through the disk because of its 
dissipation; a part goes inside the corotation, and a part makes the disk expand.

                \subsubsection{Bar pattern speed and corotation radius}
\begin{figure*}
        \centering
        \includegraphics[width=8.5cm]{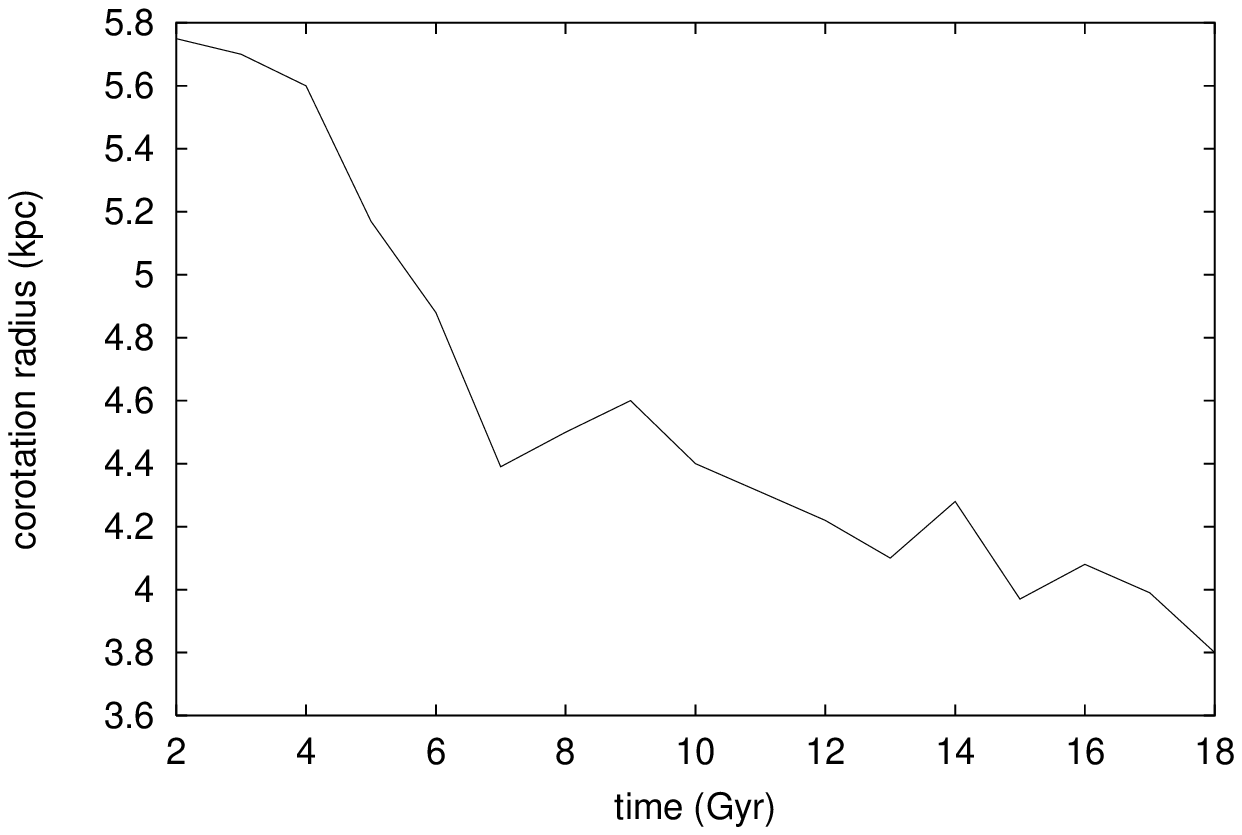}
        \includegraphics[width=8.5cm]{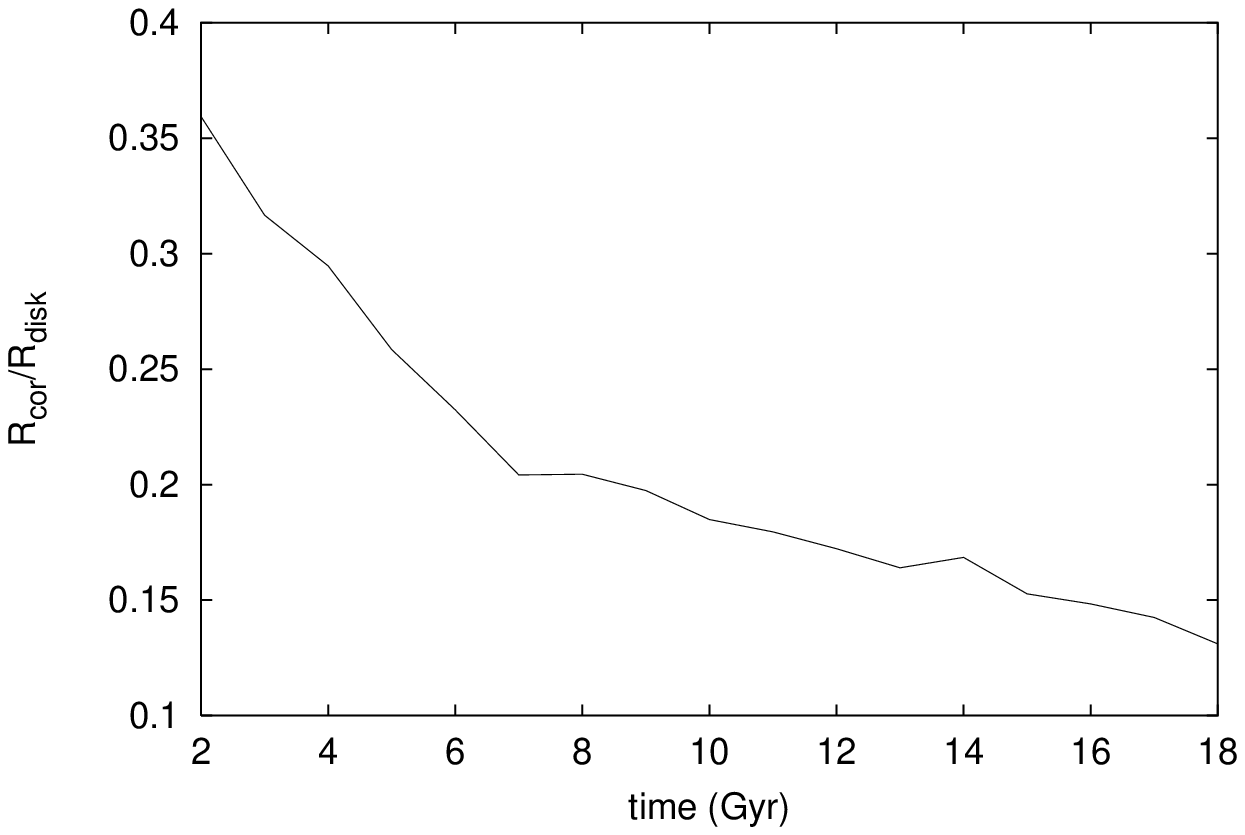}
        \caption{Corotation radius (left) and corotation to disk radius ratio (right) for model B.}
        \label{cr}
\end{figure*}

We have shown that gas accretion makes the bar pattern speed change. Yet, this does not imply that any change could be easily observed in real disks: the bar pattern speed cannot be directly observed, and the corotation radius is measured instead. So, we may wonder how the corotation radius evolves, and compare it to the disk radius. For instance for model B, the corotation radius and the corotation to disk radius ratio are given on Fig.~\ref{cr}.

The corotation radius decreases with time (see Fig.~\ref{cr}). This can be interpreted in terms of bar pattern speed: we have noticed that the pattern speed $\Omega_b$ raises with time faster than the square root of the inner mass, thus faster than the particles angular velocity $\Omega(r)$. Then, the corotation radius $\mathrm{R_{cor}}$, defined by $\Omega_b=\Omega(\mathrm{R_{cor}})$, is to decrease with time. Moreover, we know that the disk radius becomes larger, so that the corotation to disk radius ratio, which is the more relevant parameter, decreases in a significant way, as displayed on Fig.~\ref{cr}.

The bar pattern speed evolution is related to a change in the corotation radius, and makes the corotation to disk radius ratio decrease with time. Moreover, even if the bar pattern speed does not change a lot during the first Gyrs (see Fig.~\ref{bav2}), the change in the corotation radius is larger during the first Gyrs (see Fig.~\ref{cr}); then the position of the corotation is significantly modified by gas accretion even in rather young disks.

        \subsection{Gas accretion and spiral morphology}  \label{arms}

We have already briefly noticed that gas accretion rejuvenates spiral patterns. In an 
accreting disk, the main structure of the outer disk consists in spiral arms, whereas 
in non accreting disks, such a structure is rather evanescent: it is replaced by 
circular structures in about 8 Gyr.

Actually, the gas mass fraction is obviously higher; so the stellar and gaseous medium are 
globally cooler: this prevents the stability parameter $Q$ from increasing 
too much, so that the spiral structure persists in the disk. When there is no 
accretion, the first density waves heat the medium, which does not cool 
back fast enough because of lack of gas.

Moreover, the accreted gas spreading prevents the disk from getting almost empty 
of gas in specific regions
(even if does not prevent rings from appearing, as will be shown later). 
The initial gaseous density radial distribution is better preserved, except a change 
in its scale length. This let the gaseous medium stay more favourable to spiral waves 
propagation.

Rejuvenating the spiral structure does not prevent resonant ring formation.
Such rings may even be stronger. External rings are 
intensified by accreted gas accumulation in the outer disk. Inner rings, connected 
to ILR, do not seem to be affected by accretion. The most favored rings are 
those that surround the bar: they are related to the $4/1$ resonance. They are 
weaker than other rings in non accreting disks, due to gas depletion
near corotation. Accretion lets them become 
stronger. The subsequent disks are then typically RSX, SX(r), or even RSX(r) 
(X meaning A, B or AB) in the de Vaucouleurs classification. In this
classification, the meaning of SA, SAB and SB has already been
explained (see Sect.~\ref{gen}). In a SA disk, the residual bar strength, as defined in
Sect.~\ref{det} is less than 0.2, In SAB disks, the bar
strength is contained between 0.2 and 0.5 ; and in SB  disks, it is
more than 0.2 and can be much larger. We notice that for
moderate bar strengths (0.3-0.4), the disk can be either SAB or SB.
 A RSX-type disk shows an outer ring. The relations between bar strength and disk morphology have been studied by Block et al. (2001). A SX(r)-type disk
shows a ring surrounding the bar: it is not an inner ring connected
with ILR, but a larger ring. This SX(r)-type is much favored by accretion.
\bigskip

Let us try to interpret the models in terms of this classification, exactly with the distinction 
between SB and SAB galaxies. In SB disks, spiral arms generally appear in phase with the bar and 
might have the same pattern speed; in the weaker SAB ones, spiral patterns are 
sometimes separated from 
the bar, and are likely to have a lower pattern speed. 

The two disks studied in Sect.~\ref{det} can be classified SA or SAB, 
according to this definition, since  
Fig.~\ref{bav} reveals an angular speed lower for arms than for bars. However, 
the disk sometimes presents an SB aspect, for both models A and B. But the frequency 
of strong bars (SB) does not exceed  30\%, and the rate of SB disks in the 
whole galactic evolution is less than 20\%, whereas observed galaxies are divided equally 
between the three categories, thus half of barred galaxies are of SB-type.

In fact, the frequency of SB disks in simulations
depends on some input parameters. The halo to disk and 
bulge to disk ratios do not intervene there, neither does
$a_{bulge}$. The important parameters are $a_{disk}$ and $a_{halo}$. 
It is to be noted however that when  $a_{halo}$ is varied, $M_{halo}$ 
is varied also in order to keep $M_{halo}^{inner}$ constant, which is 
the relevant mass. Moreover, increasing $a_{halo}$ is equivalent to 
decrease $a_{disk}$. The runs R1a to R3c are then used here, 
and the frequency of SB disks as a function of both scale-lengths 
is given in table \ref{SBrate}. Approximatively half of barred disks 
are of SB-type, and a third of disks are SA, a third SAB, and a third 
SB, which is in good agreement with observations.

\begin{table}
        \begin{center}
        \begin{tabular}{lcc}
       \hline
       \hline
Run  &  SB to [SAB+SB] ratio & SB global rate\\
        \hline
R1a-b-c & 20 to 30 $\%$ &  15 to 20 $\%$ \\
R2a-b-c & 40 to 45 $\%$ &  25 to 30 $\%$ \\
R3a-b-c & 75 to 85 $\%$ &  50 to 55 $\%$ \\
        \hline
        \end{tabular}
        \end{center}
        \caption{Rate of SB-type disks for different parameters.}
        \label{SBrate}
\end{table}

Does accretion have some influence on the SB to SAB ratio (obviously, it reduces 
the SA rate, and raises the SAB and SB global rate, for it favours bars)?
 If we stop accretion at any epoch of simulation, the SB frequency decreases; furthermore, 
non accreting disks have lower SB  frequencies than accreting ones. So, gas accretion favours 
a high SB to SAB ratio. Accretion brings mass in the outer disk, which makes $a_{disk}$ 
rise, whereas in non accreting disks, the bar makes a heavy central condensation, which 
on the contrary reduces the disk scale-length. Finally, gas accretion favours not only 
bars, but also strong SB bars.

\section{Three-dimensional effects of gas accretion} \label{3D}

We now present the results of 3-D simulations concerning bar evolution. 
New free parameters arise, in particular the angle between the angular 
momentum of the accreted
gas and that of the disk. Also, the gas can be accreted at any radius.
For the sake of comparison with the previous 2-D-results, let us begin by in-plane accretion.

\subsection{In-plane accretion}
In these first runs, as in the 2-D-case, the gas is assumed to be accreted in the plane,
at all azimuths in the disk, with the tangential velocity corresponding to the rotation curve.
This situation corresponds to a long-term, quasi-stationary accretion over
a Hubble time, as if the optical disk 
was surrounded with slowly infalling gas, that has already settled into a larger-scale disk.
Parameters of the runs are given in table \ref{run3D1}.

\begin{table}
\begin{center}
\begin{tabular}{cc}
\hline
\hline
Run & $M_{bulge}/M_{disk}^{initial}$ \\
\hline
1 & 0.8 \\
2 & 0.6 \\
3 & 0.3 \\
\hline
\end{tabular}
\end{center}
\caption{Parameters of the 3-D runs with in-plane accretion. The value of 
$M_{halo}^{inner}/M_{disk}^{initial}$ is fixed at 0.75.
The scale-length for the disk is 5kpc, for 
the bulge 1 kpc, and for the halo 20 kpc. The accretion rate makes the disk mass 
double in 7 Gyrs, and the initial gas fraction is $20\%$. } \label{run3D1}
\end{table}

For runs 1 and 2, we observe bar destruction and re-formation. 
These runs show respectively 2 
and 3 successive bars over 15 Gyr. Run 3 presents a long-lifetime bar, that is still present in 
the disk after 13 Gyr. The bar accelerates such that its pattern speed doubles 
in about a Hubble time. The accreted gas distribution in the disk, related to bar destruction 
and re-formation in runs 1 and 2, is similar to that obtained in 2-D models. The 
vertical distribution of particles is not modified by gas accretion in these runs. 
Therefore, the results and interpretations reported for  2-D simulations are strengthened 
by these new runs.

\subsection{Out-of-plane accretion}
\begin{figure}
\resizebox{8cm}{!}{\includegraphics{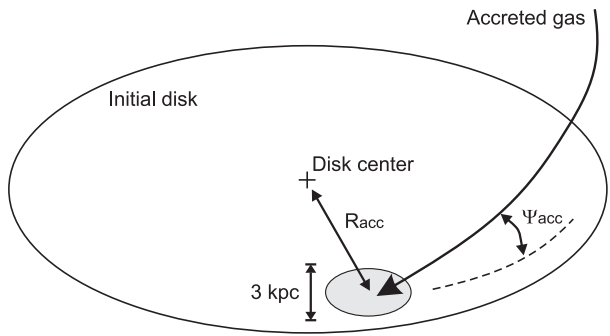}}
\caption{Model for gas accretion and parameters 
R$_{\mathrm{acc}}$ and $\psi_{\mathrm{acc}}$.}
        \label{sch}
\end{figure}

When gas is not accreted in the galaxy plane, we define R$_{\mathrm{acc}}$ the 
ratio between the accreted gas mean radius and the disk radius (see Fig.~\ref{sch}). 
The angle between the accreted gas angular momentum and the initial disk angular 
momentum is $\psi_{\mathrm{acc}}$. The accreted gas mean initial velocity has again no 
radial component. It seems reasonable to assume that 
R$_{\mathrm{acc}}$ and $\psi_{\mathrm{acc}}$ 
are nearly constant over a few Gyrs, but not over a Hubble time. 
Therefore, the accretion phase lasts only 5 Gyrs. It is started well after the first
relaxation phase, when a first bar develops in the initial disk ; a few Gyrs later, the bar
weakens, and without gas accretion, the galaxy would not be strongly barred any more.
We hereafter note $t=0$ the beginning of the accretion event, and not the 
beginning of the whole disk evolution. The accretion does not take place from all azimuths 
around the disk (at radii around R$_{\mathrm{acc}}$), but only in a 3-kpc-sized area, 
which seems more realistic for out-of-plane accretion. The mass accreted over 5 Gyrs 
is half of 
the initial disk mass. The disk evolution is followed during the accretion episode and later. 
Two important parameters, $\psi_{\mathrm{acc}}$ and R$_{\mathrm{acc}}$, are varied. 
Parameters of the run are given in table \ref{run3D2}.

\begin{table}
\begin{center}
\begin{tabular}{cccc}
\hline
\hline
Run & R$_{\mathrm{acc}}$ & $\psi_{\mathrm{acc}}$ & $M_{bulge}/M_{disk}^{initial}$ \\
\hline
REF & & no accretion  & 0.8 \\
ACC8-0 & 0.8 & 0° & 0.8 \\
ACC8-8 & 0.8 & 8° & 0.8 \\
ACC8-20 & 0.8 & 20° & 0.8 \\
ACC8-32 & 0.8 & 32° & 0.8 \\
ACC8-45 & 0.8 & 45° & 0.8 \\
ACC8-65 & 0.8 & 65° & 0.8 \\
ACC8-87 & 0.8 & 87° & 0.8 \\
ACC8-20-2 & 0.8 & 20° & 0.6 \\
ACC8-20-3 & 0.8 & 20° & 1.0 \\
ACC9-20 & 0.9 & 20° & 0.8 \\
ACC7-20 & 0.7 & 20° & 0.3 \\
ACC6-20 & 0.6 & 20° & 0.3 \\
ACC3-25 & 0.3 & 20° & 0.8 \\
\hline
\end{tabular}
\end{center}
\caption{Parameters varied for out-of-plane accretion. 
R$_{\mathrm{acc}}$ is the ratio between the accreted gas radius and the disk radius, 
and $\psi_{\mathrm{acc}}$ is the angle between the angular momenta of the 
accreted gas and the initial disk, as described on Fig.~\ref{sch}. We choose $M_{halo}^{inner}/M_{disk}^{initial}=0.75$.}
\label{run3D2}
\end{table}

\subsubsection{Large-radius accretion}
In runs ACC8-0 to ACC8-87, the gas is accreted in the outer disk with several different 
values of R$_{\mathrm{acc}}$. The gas accretion always re-inforces the bar significantly.
Warps are also formed in the outer disks, while the initial disk axis rotates.

	\paragraph{General disk evolution}
When gas is accreted with an angular momentum inclined to that of the disk, it is submitted to
restoring forces from the disk, that make the orbits precess. The precession is differential,
and the various gaseous orbits, at different radii will soon intersect, and dissipation will
occur and make newly accreted gas align to the disk plane.
Therefore all the matter will gather in a disk, which axis will 
slowly change orientation in a few 
Gyrs (see Jiang \& Binney, 1999). This change in the disk orientation is observed in the present
simulations (see Fig.~\ref{profil}). There is also a flaring in the re-oriented gaseous disk, 
while this flaring disappears during the change of orientation. 
A polar ring also appears in run ACC8-87.

\begin{figure}
        \centering
%\resizebox{8cm}{!}{\includegraphics{flare.png}}\\
%\resizebox{8cm}{!}{\includegraphics{flare2.png}}
        \caption{Edge-on views of the gaseous component at t=6 Gyr 
(1 Gyr after the end of accretion) 
for run ACC8-45. The view axis is perpendicular to the major axis of the bar (top), 
then parallel to 
it (bottom). For this run, the change in disk orientation is almost over at t=6 Gyr.}
        \label{profil}
\end{figure}

	\paragraph{Bar strength and out-of-plane accretion}
For runs ACC8-8 to ACC8-87, a bar appears in the disk. It is very long around t=3-5 Gyr, 
as shown on Fig.~\ref{sbf} for run ACC8-45 at t=5 Gyr. For this runs, the maximal bar strength 
is 0.75, and exceeds 0.85 for runs ACC8-0, 8 and 20. In the non-accreting disk REF, the primordial 
bar is weakened at $t=0$ and the bar strength is never more than 0.3. Bars re-formed by 
out-of-plane gas accretion are much stronger than the bar re-formed in the in-plane accretion 
runs, where the bar strength never exceeded 0.5-0.6 in either 2-D or 3-D runs, 
whatever the different 
parameters were. The lower strength of previous bars are obvious when comparing Figs.~\ref{sbf} 
and \ref{disk2}.

\begin{figure}
%\resizebox{8cm}{!}{\includegraphics{sbf.png}}
\caption{Stellar and gaseous particles viewed perpendicular to the
  re-oriented disk at t= 5 Gyr, i.e. the end of the
  accretion event, for run ACC8-45. This disk shows two spiral arms in
  phase with a very strong bar.}
        \label{sbf}
\end{figure}

Thus the present out-of-plane gas accretion model appears more favourable to
barred waves than the previous in-plane accretion models. To interpret this, 
let us consider the run ACC8-0: this disk also develops a very strong bar, 
of maximal strength 0.9, while the angle of accretion is zero. In fact,
the common new feature introduced in the 3-D runs is to limit
the accretion to a 3-kpc-sized area of the disk. This
increases the non-axisymmetric component of the
gravitational potential and favors a stronger bar. Moreover, when $\psi_{\mathrm{acc}}>0$, the accretion paths
intersect the disk plane around two points: this strengthens $m=2$ harmonic of the
potential, which may explain the enhancement of two spiral arms
in the disks.

\begin{figure}
\resizebox{8cm}{!}{\includegraphics{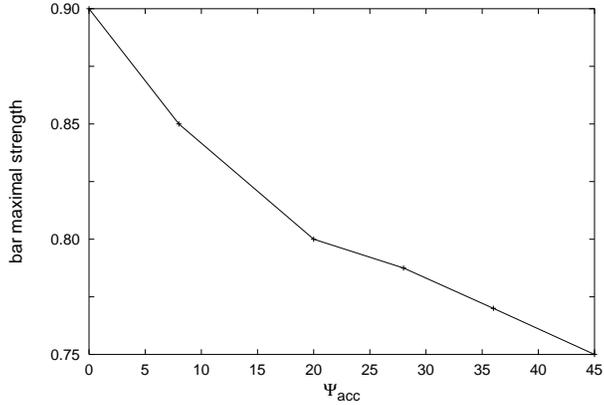}}
\caption{Maximal bar strength over the whole bar evolution, as a
  function of $\psi_{\mathrm{acc}}$, with R$_{\mathrm{acc}}$=0.8.}
        \label{bar_psi}
\end{figure}

In summary two phenomena may explain why gas accretion favor bars in
disk galaxies:
\begin{itemize}
\item the supply of mass to the disk, which reduces the relative mass
  of the central stabilizing condensation.
\item the asymmetric distribution of the accreted gas.
\end{itemize}
It is difficult to disentangle the dependence of each of them on the
accretion parameter $\psi_{\mathrm{acc}}$, since both phenomena occur
simultaneously. Yet, the main trends are:
\begin{itemize}
\item with larger $\psi_{\mathrm{acc}}$, the gas accretion onto the disk
is longer and slower, which does not favor bar formation.
\item with larger $\psi_{\mathrm{acc}}$, the  distribution of the accreted 
gas is more asymmetric, and the $m=2$ harmonic of the resulting
  potential is higher.
\end{itemize}
The two mechanisms to form, re-form, or
strengthen a bar, have thus opposite dependence on
$\psi_{\mathrm{acc}}$. The evolution of the actual bar
strength on $\psi_{\mathrm{acc}}$ will allow us to determine which is
prevailing.
The bar maximal strength as a function of
$\psi_{\mathrm{acc}}$ for runs ACC8-0 to 45 is plotted on Fig.~\ref{bar_psi}.

We find that the maximal bar strength decreases when
$\psi_{\mathrm{acc}}$ increases. 
It appears that the influence of a strong and rapid mass supply
dominates the process of bar re-formation.

        \paragraph{Lifetime of the bar and gas accretion}

\begin{figure}
        \centering
\resizebox{8cm}{!}{\includegraphics{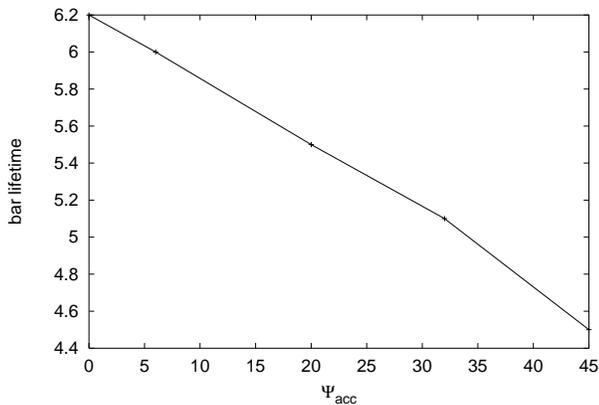}}
        \caption{Lifetime of the bar as a function of the accretion parameter
          $\psi_{\mathrm{acc}}$. The bar episode ends
          when the bar strength is less than 0.2. The mean
          accretion radius is R$_{\mathrm{acc}}$=0.8}
        \label{lft}
\end{figure}

Let us now examine the dependence of the bar lifetime on
$\psi_{\mathrm{acc}}$. Considering that the bar is destroyed when its strength 
drops below 0.2, the lifetime of the bar in each run is shown on
Fig.~\ref{lft}. The first point is that all these bars are rather short-lived
features. This is not surprising, for at least two reasons: first the gas
accretion is supposed to be limited in time, and 
occur during only a few Gyrs; second the formed bars are very strong, 
so the induced mass transfers destroy them
in only a few Gyrs. 
The second point is the dependence of the bar lifetime
with the accretion angle. The bar lifetime is longer for
 smaller $\psi_{\mathrm{acc}}$, and therefore the
strongest bars have here the longest lifetimes. It is
well-known that in non-accreting disks, the strongest bars are the most
evanescent ; thus the bar lifetime is strongly influenced by gas accretion. 
A low $\psi_{\mathrm{acc}}$ favors the mass supply to bars,
which lengthens their lifetime.
Accretion not only creates a bar, but also maintains it. 

        \paragraph{Evolution of the bar pattern speed}

\begin{figure}
        \centering
\resizebox{8cm}{!}{\includegraphics{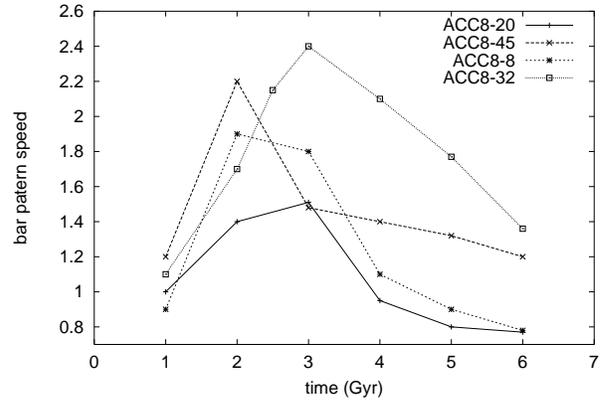}}
        \caption{Bar pattern speed as a function of time for several runs.}
        \label{omeg}
\end{figure}

The evolution of the bar pattern speed with time is shown for different
runs on Fig.~\ref{omeg}. Two steps are clearly seen:
when the bar is rather young (i.e. before 2-3 Gyrs), the pattern speed
rises. Then, after a peaked maximum, it decreases while the bar is 
destroyed. Without accretion, the
bar rotation would always slow down (Combes \& Sanders, 1981) ; the
acceleration observed here at the beginning is related to gas accretion. This
succession of acceleration and deceleration of the bar rotation is
similar to that observed for 2-D models with bar destruction and
re-formation (see Fig.~\ref{bav2}).

In summary, out-of-plane gas accretion, like in-plane accretion, forces the bar pattern
speed to increase, while bars naturally tend to slow down.

\subsubsection{Small-radius accretion and dependence on accretion radius}
When gas is accreted from outside the plane, the choice of accretion radius
R$_{\mathrm{acc}}$ is free. In the simulation ACC3-25, the accretion radius is
chosen very small. The disk radius decreases in a few Gyrs, and no bar is
formed in the disk. However such a situation is rather unusual, since gas 
arriving at larger radii will be accelerated in the infall, and the accretion
will take a long time. A rapid deceleration requires violent conditions
and strong dissipation, that occur in galaxy collisions and
mergers. A very small radius of accretion with high mass accretion rate is 
not likely to resemble a quasi-steady accretion.

Let us now study the influence of the accretion radius when
R$_{\mathrm{acc}}$ is of the same order as $r_{disk}$: the runs involved
are ACC6-20, 7-20, 8-20 and 9-20. There is no
significant dependence of the bar evolution on R$_{\mathrm{acc}}$. In
fact, in the above runs, a weakened lens-shaped bar is present in
the disk when we start accreting gas: its corotation radius is 7
kpc. Then, for runs ACC6-20, 7-20, 8-20 and 9-20, the gas is accreted
outside the corotation (the accretion radius is more than 7.5 kpc in
those runs). As we know from 2-D studies that the important factor is
the relative position of the accreted gas with respect to corotation
radius, we can understand that the accretion radius does not influence
the bar evolution, provided that this accretion radius is larger than 
the corotation one.

\subsection{General situations}
There are many free parameters concerning the way gas is accreted on a galactic disk. 
We have not studied every possible situation, yet we have presented several 
models that enables us to describe more general situations. A real disk is thought to 
accrete matter both from its plane and outside it, with parameters varying with time. 
Thus, a real accretion is likely to be a succession of elementary events, each of them being 
represented by one of our models. To illustrate this, we now present the evolution of a 
disk that accretes gas in its plane over a Hubble time, and from out of its plane 
during 5 Gyrs at the middle of its life.

In this run, the initial bulge to disk mass ratio is 0.7, and the halo to disk is 0.75. 
In-plane accretion is implemented as before with a rate that makes the initial disk mass 
double in 10 Gyrs. From $t=6$ Gyr to $t=11$ Gyr, out-of-plane accretion takes place with 
$R_{acc}=0.8$, $\Psi_{acc}=25°$, and a mass flow corresponding to the initial disk mass in 8 Gyrs.

Before $t=6$ Gyr, the bar evolution is rather similar to the 2-D run A. Here, the initial 
bar is very weakened at $t=5$ Gyr, while its strength had reached 0.68. Then, a 
new bar appears in the disk around $t=6$ Gyr (see Fig.~\ref{disk3}). This bar is initially 
rather weak, then out-of-plane accretion strengthens it, as it has already been shown. 
The maximal bar strength is 0.92. In the out-of-plane runs with a maximal bar strength 
larger than 0.8, the bar strength drops below 0.4 in about 2 Gyrs. Yet, the very strong bar 
formed by out-of-plane accretion is here maintained by in-plane accretion. For instance, 
at $t=10.5$ Gyr, the bar has the same aspect as at $t=8$ Gyr, as shown on Fig.~\ref{disk3}, 
and is bar strength still is 0.85. Later, a bar whose strength is larger than 0.34 
remains until the end of the simulation, at $t=15$ Gyr. 

\begin{figure*}
        \centering
%\resizebox{5.9cm}{!}{\includegraphics{d6.png}}
%\resizebox{5.9cm}{!}{\includegraphics{d8.png}}
%\resizebox{5.9cm}{!}{\includegraphics{d10.png}}
        \caption{Stellar and gaseous particles viewed perpendicular to the disk 
plane at $t=$6, 8, and 10.5 Gyr, for the in- and out-of-plane accretion run.}
        \label{disk3}
\end{figure*}

Concerning the evolution of the bar pattern speed (normalized to 1 at $t=0$), the second bar formed 
around $t=6$ Gyr is faster than the initial one (1.36), and the pattern speed is triggered 
by the out-of-plane event around $t=8$ Gyr (2.35). After being very strong, the bar first slows 
down (1.97), but is then re-accelerated by the maintenance effect, while it becomes shorter, 
as in the 2-D run B: the bar pattern speed at $t=15$ Gyr is finally 2.31. Thus, the
bar pattern speed doubles or even more over a Hubble time, as was the case for runs A and B 
(Sect.~\ref{det}).

\section{ Discussion of numerical models} \label{discu}
General discussions are given in subsections \ref{dis2d} and \ref{dis3d}. A discussion of the ISM scheme is given in subsection \ref{gasdyn}.

\subsection{2-D models} \label{dis2d}
It is normal to expect that gas accreting galaxies maintain a 
relatively high gas fraction in their disk. However,
even if the disk mass doubles in a few Gyrs, the gas to total (stars +gas) 
mass ratio is generally not more than 0.5: the star formation and gas
consumption rate is of the same order as the accretion rate. The
long-term evolution of a gas accreting disk depends not only on the
dynamics of this gas, but also on the evolution of stars formed in
accreted matter, particularly for the consequences
on stellar bars. Mass exchanges between stellar and gaseous phases have
to be treated realistically. Our simulations thus include detailed
star formation and stellar mass-loss schemes.
In particular, the stellar mass-loss is
not approximated as instantaneous: this is rather important, since
mass re-injection over several Gyrs would change radial flows of
matter, which then influence the bar properties (see Jungwiert et
al. 2001). However, there are still some shortcomings in the present model,
since gas is assumed isothermal and the dependence of the yield
on metallicity -that evolves with time,
and also with accretion- is not included; it is hoped that these are only minor
effects, that will not significantly affect the mass accretion rate.

The dissipation time-scale $k$ for the gas dynamics cannot be exactly justified, even if a realistic value is of a few Gyrs. Yet, this adjustable value has no major influence on the morphology of disks. When this time-scale $k$ is varied between $5$ and $10$ Gyrs, 
the long-term evolution of the disk remains unchanged: the
gas reaches an equilibrium between dissipation and heating by density
waves which is not significantly influenced by the value of $k$.

\subsection{3-D models} \label{dis3d}
Our three-dimensional simulations include all the major phenomena
that influence the properties of bars: gas dynamics, star formation,
and stellar mass-loss and feedback. 
However, the resolution of our 3-D models is lower than in 2-D:
the cell dimension is 450 pc (instead of 100 pc).
The dark matter is represented by a rigid potential, and
there are still many physical processes that are simplified or ignored
(the multiphase nature of the gas, for instance). 
May be the most relevant uncertain part is the rate of star formation,
that might not be a Schmidt law.

\subsection{Gas dynamics}\label{gasdyn}
Different schemes were used to treat the gas dynamics in our simulations. They have given comparable results. They do not include a precise description of the real ISM. A more precise study may treat gathering and fragmentizing of gas clouds, and energy feedback from star formation. However, a simplified description of the ISM seems to be suitable for a first approach of large-scale accretion. Actually, what may be more questionable is the dissipative scheme, for it does not conserve exactly angular momentum. Sticky-particles codes are generally thought to be rather convenient ; yet, we study gas-rich galaxies, which makes such a shortcoming be suspect. We have shown that angular momentum transfers and related mass exchanges are here fundamental processes, so we have to verify that the local non-conservation of the angular momentum does not invalidate our results. This is done in appendix \ref{append}.

\section{Conclusion}

The numerical simulations presented in this work reveal how different
can be the evolution of spiral galaxies with and without important
gas accretion. 
While an isolated galaxy may spontaneously form a bar, consume its
gas in stars, and when the stellar component is hot enough, see its bar
weaken to a lens, and evolve definitely to an early-type system,
a galaxy with significant gas accretion, can reverse this evolution,
and experience several new bar episodes.
The detailed processes are controlled by the strength of the bar and
its gravity torques. The accreted gas may stay for a while in the outer parts
of the disk, prevented to flow in by a strong bar; as soon as the bar
weakens, the gas will progressively populate the disk, and form new stars
there. This increases the self-gravity of the disk with respect to the
stabilizing effect of the central bulge,  favoring a new bar instability.
Under the influence of this new bar, the material inside its corotation
is driven inwards, and through vertical resonance, thickens to enrich the
bulge again. This regulates the bar strength. 

In the course of the simulations, three to four bar episodes have 
been followed. An interesting discovery is that the pattern speed
of the bar is increasing form one bar to the next. This can be
considered as a rejuvenating process. Indeed, the normal evolution
of a bar is to slow down, while trapping more and more material,
and lengthening, so that its corotation radius increases (a bar cannot
extend beyond its corotation, since periodic orbits are perpendicular
here). 
When another bar forms, it is shorter than the previous one.
Its pattern speed is higher, not only because there is then more mass
contained in a given radius, due to mass accretion. The pattern speed
is even higher than what would be expected from
the square root of mass increase.
This increase may be related to the fact that mass is more and more 
concentrated during this evolution, and the galaxy is shifting 
progressively to early-types, with massive bulges. The bar is then 
shorter with respect to the galaxy radius (as shown by Combes \& Elmegreen
1993). Also, gas accretion has the effect of increasing the inner
angular momentum per unit mass. The bar is shorter and rounder.
 
In the course of evolution with gas accretion, the radius of the
galaxy expands, a phenomenon also controlled by the bar strength,
which gravity torques can maintain or expel the gas outwards.
One of the most important consequence of gas accretion, is to maintain
an almost permanent (though recurrent) spiral structure through the
disk, and also a significant star formation rate along a Hubble time.
This solves the problem of explaining the almost constant star formation
rate of spiral galaxies along their history (e.g. Kennicutt 1983),
and also the presence until today of conspicuous spiral structure 
in most spiral disks. 

The possibility to accrete gas from a direction highly inclined
with respect to the galaxy plane introduces new processes that can
be also favorable to bar reformation. 
The gas may not have to wait for a weakening of the previous bar to
flow in, and the distribution is then asymmetric, with an m=2
extra contribution, which makes bars become much stronger.

One of the consequences of the evolution described in this work is that the
bar pattern speed can no longer give the evolution state or "age" of a barred
galaxy. Significant gas accretion can reverse the evolution and 
increase the pattern speed. Also, the argument that dynamical friction
would have slowed down the bars along their evolution would have
to be revised (e.g. Debattista \& Sellwood 1998).
Another consequence is that the morphology of a galaxy,
according to its environment, and the availability to accrete
cold gas, may oscillate from late to early-types three or four times,
before inexorably shift to an early-type galaxy.

The maintenance of conspicuous spiral structure in most disk galaxies
may mean that galaxies have sufficient gas to accrete in their outer
parts, to constantly renew disk instability, even in the presence
of a heating stellar disk and increasing bulge mass.
The relative absence of bars in galaxies at high redshift (e.g.
van den Bergh et al. 1996) could be interpreted as a very short 
life-time of the bar episode, in the presence of dominating gas fraction
in very young galaxies. Indeed, the fraction of barred galaxies at a given
epoch is interpreted, in this recurrent bar scenario, as the relative
time spent by a galaxy in the bar episode relative to the un-barred one.

The present work has only considered general properties of gas accreting
spiral galaxies, but many parameters (mass distribution, morphological
types, geometry of the accretion, etc..) have to be explored before
all consequences of accretion on galaxies can be drawn, which 
will be done in future work.

\begin{appendix}
\section{Numerical schemes for gas dynamics and non conservation of integrals of motion}\label{append}  

\subsection{Variations of angular momentum and energy}

\begin{table}
\centering
\begin{tabular}{lrc}
\hline
\hline
 & Integral of motion & Value after 5 Gyrs \\
\hline
\multicolumn{2}{l}{2-D viscosity scheme} & \\
& \hspace{1cm}Angular momentum & 0.97 \\
& \hspace{1cm}Kinetic energy & 0.60 \\
\multicolumn{2}{l}{3-D sticky-particles scheme} &\\
& \hspace{1cm}Angular momentum & 0.98 \\
& \hspace{1cm}Kinetic energy & 0.53 \\
\hline
\end{tabular}
\caption{Relative variations of angular momentum and energy of gas particles over 5 Gyrs, for the control run without accretion, star formation, and stellar mass-loss. The initial values are normalized to 1.} \label{app_tab}
\end{table}
We have shown that redistribution of angular momentum inside the disk by spirals and bars waves was partly responsible for an increase in the bar pattern speed, which is an important consequence of gas accretion. However, the numerical codes we have used do not conserve angular momentum. Both ISM models, viscosity scheme or sticky-particles code, dissipate the angular momentum of gas clouds, at least with the chosen parameters. This could be a real process if a spin was stored in gas particles, but it is not the case in our simulations. Thus, we discuss in this appendix the non-conservations of angular momentum by gas dynamics schemes, in relation with our results, and more generally in all N-body simulations. Another basic integral, the energy, is not conserved in the ISM: here it corresponds to a physical process, for gas clouds radiate away a significant energy.

In a sticky-particles code, when two gas particles collide, their local angular momentum is generally not conserved. According to equations \ref{stick1} to \ref{stick2}, the angular momentum of the particles in the referential of their mass center before the collision is:
\begin{equation}
\vec{L}=(\vec{r}_2-\vec{r}_1)\times\vec{v}_t\frac{m_2 m_1}{M}
\end{equation}
After the collision, only velocities have changed, and the angular momentum is:
\begin{equation}
\vec{L'}=(\vec{r}_2-\vec{r}_1)\times\vec{v'}_t\frac{m_2 m_1}{M}
\end{equation}
\begin{equation}
\vec{L'}=\beta_t\vec{L}
\end{equation}
Thus, the sticky-particles code conserves the angular momentum only when $\beta_t=1$. It does not when $\beta_t=0.65$. However, during a collision, the angular momentum around the mass center is changed, but the momentum of the mass center around the galactic center remains unchanged. Actually, the second one is much larger: its mean value, for each pair of gas particles susceptible to collide, is almost $10^3$ larger than the mean value of the first one (this measure is for run 0F1). Thus, only a small fraction of the angular momentum can be reduced. We are now to show quantitatively that the variations of angular momentum induced by both dissipative schemes are not as important as the effects of gas accretion that we have shown.

Accretion is responsible for changes in angular momentum and energy, thus, in order to quantify the non conservation of integral of motions due to our code, we compute the variation of energy and angular momentum in an isolated disk (run 0F1 in Sect. \ref{sec3}). Star formation and stellar mass-loss modify the gas content and the importance of gas dynamics, thus we also suppress them. In table \ref{app_tab} we give the relative variation of angular momentum and energy in the gaseous disk over 5 Gyrs. The quantities taken away by particles going out of the grid have been accounted for, thus the gas dissipation is the only cause for the observed variations. The ISM undergoes an important loss of energy, that is physically justified, and a small loss of angular momentum: as explained qualitatively before, collisions in the 3-D code, and viscosity in the 2-D one, dissipate only a small fraction of the angular momentum, that cannot be compared with the large supply of angular momentum by accretion.

\subsection{Influence of the angular momentum conservation on previous results}
We have explained that numerical schemes for gas dynamics do not conserve exactly the angular momentum. We wish to show that this is not a critical point, and does not weak our conclusions about gas accretion.

The dissipative schemes we have used suppress local vortices, and tend to reduce the angular momentum of two colliding particles. This may modify the dynamics of the galactic center. Thus, we modify the value of $\beta_t$ to 1, then the angular momentum is conserved. We test simulations with this parameter against previous results established with a sticky-particles code, i.e. 3-D results. For 2-D results, we compare the viscosity scheme to a modified version in which only the radial velocities are reduced, still in order to conserve angular momentum: in this modified version, a particle at a position $\vec{r}$ from the disk center, with a velocity $\vec{v}$, in a cell in which the mean velocity is $<\vec{v}>$, will undergo a force that reads:
\begin{equation}
\vec{f}=-k\left( ( \vec{v} - <\vec{v}> ) \cdot\vec{r} \right) \frac{\vec{r}}{r^2}
\end{equation}
We also compare the 2-D viscosity scheme with a 2-D sticky-particles code, using $\beta_t=1$.

\begin{figure}
        \centering
\resizebox{8.5cm}{!}{\includegraphics{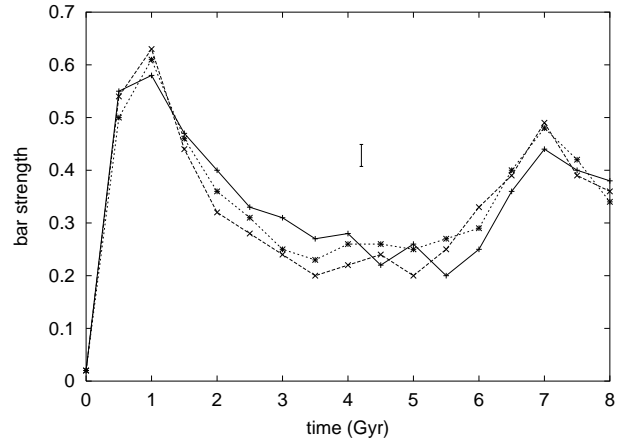}}
        \caption{Bar strength in a 2-D simulation. Solid line: angular momentum is conserved (sticky-particles $\beta_t=1$), dashed line: angular momentum is not conserved (viscosity scheme), dotted line: angular momentum is conserved (modified viscosity scheme). We show the scatter (error bar) measured on 5 simulations starting with 5 different, randomly chosen, initial position of the particles, all the physical parameters remaining unchanged.}
        \label{test1}
\end{figure}
\begin{figure}
        \centering
\resizebox{8.5cm}{!}{\includegraphics{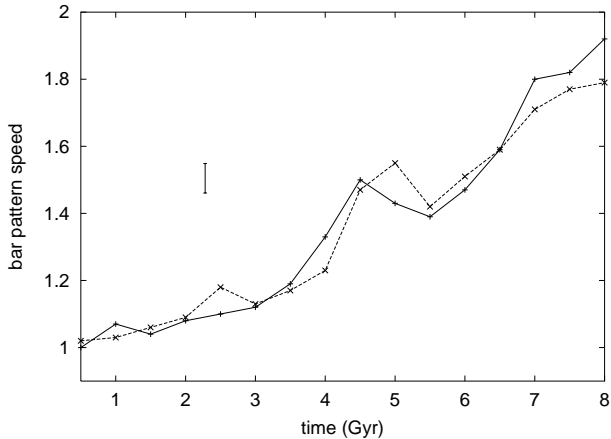}}
        \caption{Bar pattern speed in a 3-D simulation. Solid line: angular momentum is conserved ($\beta_t=1$), dashed line: angular momentum is not conserved ($\beta_t=0.65$). The error bar has the same meaning as on figure \ref{test1}.}
        \label{test2}
\end{figure}

On Fig.~\ref{test1}, we compare the evolution of the bar strength in a 2-D disk for the viscosity scheme, its modified version that conserves angular momentum, and the sticky-particles code with $\beta_t=1$. Fig.~\ref{test2} shows the evolution of the bar pattern speed in a 3-D run, for the sticky-particles code with $\beta_t=0.65$ and $1$. The results are quite similar for the different codes, and the differences between each of them is not significant: they do not largely exceed the scatters obtained on 5 simulations starting with the same values for physical parameters, but with 5 different, randomly chosen, initial positions for the particles.

For both 2-D and 3-D models, we conclude that conserving the local angular momentum or not does not significantly influence the bar evolution and the spiral structure. This confirms all our previous results. More generally, this shows that the 2-D linear dissipative scheme and the sticky-particles code are rather convenient when one studies the evolution of disk galaxies. Furthermore, the dissipation in the real ISM is likely to be somewhere in between our two schemes. One scheme exactly conserves the angular momentum, the other one annihilates local vortices: real gas clouds may dissipate a fraction of angular momentum in cloud-cloud collisions, and store it in their spin. As the results are similar with or without dissipating angular momentum, it seems likely that the response of the real ISM to gas accretion on galactic disk is similar to what we have drawn out from our simulations.

\end{appendix}

\begin{acknowledgements}
The 3-D computations in this work have been realized on the Fujitsu
NEC-SX5 of the CNRS computing center, at IDRIS. The 2-D simulations
have been carried out on a PC desktop computer.
\end{acknowledgements}

\end{document}